\documentclass[11pt,a4paper]{article}
\usepackage[a4paper,margin=1in,footskip=0.25in]{geometry}
\usepackage{amsthm}
\usepackage{amsmath}
\usepackage{amssymb}
\usepackage[dvipdfmx,unicode,bookmarks=true, bookmarksnumbered=true, bookmarksopen = false, colorlinks=true]{hyperref}
\usepackage{graphics}
\usepackage{epsfig}
\usepackage{geometry}
\makeatletter \renewenvironment{proof}[1][\proofname]
{\par\pushQED{\qed}\normalfont\topsep6\p@\@plus6\p@\relax\trivlist\item[\hskip\labelsep\bfseries#1\@addpunct{.}]\ignorespaces}{\popQED\endtrivlist\@endpefalse} \makeatother

\theoremstyle{plain}
\newtheorem{thm}{Theorem}
\newtheorem{lem}{Lemma}
\newtheorem{defn}{Definition}
\newtheorem{exmp}{Example}

\newtheorem{cor}{Corollary}
\newtheorem{prop}{Proposition}

\begin{document}
\author{Yangbo Song\thanks{Department of Economics, UCLA. Email: darcy07@ucla.edu.}}
\date{\today}
\title{Social Learning with Endogenous Network Formation\thanks{I am grateful to Ichiro Obara, William Zame, Sushil Bikhchandani, Simon Board, Moritz Meyer-ter-Vehn, Marek Pycia, Navin Kartik, and a number of seminar audiences for suggestions which have improved the paper.}}
\maketitle

\begin{abstract}

I study the problem of social learning in a model where agents move sequentially. Each agent receives a private signal about the underlying state of the world, observes the past actions in a neighborhood of individuals, and chooses her action attempting to match the true state. Existing literature shows that with unbounded private beliefs, asymptotic learning occurs if and only if agents observe a close predecessor. However, a prevailing assumption in these studies is that the observation structure is exogenous. In contrast to most of the previous literature, I assume in this paper that observation is endogenous and costly. More specifically, each agent must pay a cost to make any observation and can strategically choose the set of actions to observe privately. I introduce the notion of maximal learning (relative to cost) as a natural extension of asymptotic learning: society achieves maximal learning when agents can learn the true state with probability 1 in the limit after paying the cost. I show that observing only a close predecessor is no longer sufficient for learning the true state with unbounded private beliefs and positive costs. Instead, maximal learning occurs if and only if the size of the observations extends to infinity.

\begin{flushleft}
\textbf{Keywords:} Information aggregation, Social earning, Network formation, Herding, Information cascade, Information acquisition
\end{flushleft}

\begin{flushleft}
\textbf{JEL Classification:} A14, C72, D62, D83, D85
\end{flushleft}
\end{abstract}

\newpage
\section{Introduction}

How do people aggregate dispersed information? Imagine a scenario with a large number of agents, each trying to match her action with some underlying state of the world, e.g., consumers choosing the highest quality product, firms implementing the technology with the highest productivity, etc. On one hand, each agent may have some informative but noisy private signal about the particular state; combining all the signals will yield sufficient information for the entire society to learn the true state. However, such private signals are typically not directly observable to others; in other words, information is decentralized. On the other hand, an agent's action is observable and informative regarding her knowledge; thus, by observing one another, agents can still hope for some level of information aggregation. Therefore, it is of great importance to investigate the relation between the type of observation structure and the type of information aggregation that is achievable.

A large and growing literature has studied this problem of learning via observation. Renowned early research, such as Bikhchandani, Hirshleifer and Welch\cite{BHW}, Banerjee\cite{Banerjee} and Smith and Sorensen\cite{SS}, demonstrate that efficient information aggregation may fail in a perfect Bayesian equilibrium of a dynamic game, i.e., when agents act sequentially and observe the actions of all their predecessors, they may eventually ``herd'' on the wrong action. In a more recent paper, Acemoglu et al.\cite{ADLO} consider a more general, and stochastic observation structure. They point out that society's learning of the true state depends on two factors: the possibility of arbitrarily strong private signals (unbounded private beliefs), and the nonexistence of excessively influential individuals (expanding observations). Expanding observations refer to the condition that in the limit, each agent observes the action of some predecessor whose position in the decision sequence is not particularly far from her own. In particular, Acemoglu et al.\cite{ADLO} show that when private beliefs are unbounded, a necessary and sufficient condition for agents to undertake the correct action almost certainly in a large society is expanding observations.

In the studies discussed above and many other related works, a common modeling assumption is that the network of observation is exogenous: agents are not able to choose whose actions to observe or whether to observe at all. In practice, however, observation is typically both costly and strategic. First, time and resources are required to obtain information regarding others' actions. Second, an agent would naturally choose to observe what are presumably more informative actions based on the positions of individuals in the decision sequence. In this paper, I analyze an endogenous network formation framework where the choice of observation depends on the private signal, and address how it affects the aggregation of equilibrium information.

The outline of the model can be illustrated with the following example. Consider a firm facing the choice between two new technologies, and the productivity of these technologies cannot be perfectly determined until they are implemented. The firm has two sources of information to help guide its decision as to which technology to implement: privately received news regarding the productivity of the two technologies, on the one hand, and observing other firms' choices, on the other. The firm knows the approximate timing of those choices, but there is no direct communication: it is unable to obtain the private information of others and can know only which technology they have chosen. Moreover, observation is costly and is also a part of the firm's decision problem -- the firm must decide whether to make an investment to set up a survey group or hire an outside agent to investigate other firms' choices. If it chooses to engage in observation, the firm must decide which of the other firms it would like to observe because there is likely a constraint on how much information can be gathered within a limited time and with limited resources. A similar example would be the choice of a consumer between two new products (e.g. two new smartphones). In this case, the consumer cares about which product has higher quality, and faces the problem of whether to spend time and effort (e.g. on an online forum) collecting information about other consumers' choices before she makes her own.

More formally, there is an underlying state of the world in the model, which is binary in value. A large number of agents sequentially choose between two actions with the goal of matching their action with the true state. Each agent receives a private signal regarding what the true state is, but the signal is not perfectly revealing. In addition, after receiving her signal, each agent can pay a cost to observe a number of her predecessors, i.e., to connect with a certain neighborhood\footnote{For the main part of the paper, the observation cost is assumed to be fixed. The more general case where the cost depends on the number of observations is discussed in Section 6.}. Exactly which of the predecessors to observe is the agent's strategic choice, and the number of others to observe is limited by an exogenous capacity structure. By observing a predecessor, the agent knows the action of the other, but not the other's private signal or which agents that have been observed by the other\footnote{If observing an agent also reveals her observation, there exists information diffusion in the game. In the present paper, I discuss this case after presenting the main results.}. After this process of information gathering, the agent makes her own choice.

In the present paper, I address the central question in this line of research under the new context of endogenous network formation, i.e., when can agents achieve the highest possible level of learning (taking the right action)? In the literature, this scenario is referred to as \textit{asymptotic learning}, which means that the true state is revealed in the limit, and that information aggregation in equilibrium would be the same as if all private information were public. When observation is endogenous, asymptotic learning may never occur in any equilibrium (e.g., when the cost of observation is too high for a rational agent to acquire information). Hence, asymptotic learning no longer characterizes the upper bound of social learning with endogenous observation. I therefore generalize the notion of the highest equilibrium learning probability to \textit{maximal learning}, which means that, in the limit, information aggregation in equilibrium would be the same as it would be if an agent could pay to access and observe all prior private information. In fact, maximal learning reduces to asymptotic learning when the cost of observation is $0$ or when private beliefs are relatively weak with respect to cost.

There are thus two central factors determining the type of learning achievable in equilibrium. The first is the relative precision of the private signal, which is represented by the relation between the likelihood ratio and the cost of observation. Consider a hypothetical scenario in which an agent can pay to directly observe the true state. If a private signal indicates that the costly acquisition of the true state is not worthwhile, then the agents have strong private beliefs; otherwise, they have weak private beliefs. The extreme case of strong private belief is unbounded private belief, i.e., the likelihood ratio may approach infinity and is not bounded away from zero. If the likelihood ratio is always finite and bounded away from zero, then the agents have bounded private beliefs. Note that bounded private belief can also be strong, depending on the cost. The second key factor is the capacity structure, which describes the maximum number of observations for each agent. I say that the capacity structure has infinite observations when the number of observations goes to infinity as the size of the society becomes arbitrarily large; otherwise, the capacity structure has finite observations. Infinite observations imply that the influence of any one agent's action on the others becomes trivial as the size of the society grows because that action only accounts for an arbitrarily small part of the observed neighborhood.

The main results of this paper are presented in two theorems. Theorem 1 posits that when the cost of observation is zero and agents have unbounded private beliefs, asymptotic learning occurs in every equilibrium. As discussed above, the previous literature has shown that a necessary and sufficient condition for asymptotic learning under unbounded private beliefs is expanding observations. This theorem implies that when observation can be strategically chosen with zero cost, the condition of expanding observations becomes a property that is automatically satisfied in every equilibrium, i.e., every rational agent will choose to observe at least some action of a close predecessor. This can be regarded as a micro-foundation for the prevalence of expanding observations when observation is free.

Theorem 2 is this paper's most substantive contribution and demonstrates that a sufficient and necessary condition for maximal learning is infinite observations when cost is positive and private beliefs are strong. Multiple implications can be drawn from this result. First, when cost is positive and private beliefs are strong, asymptotic learning is impossible because there is always a positive probability that an agent chooses not to observe. In other words, maximal learning marks the upper bound of social learning. Second, to achieve maximal learning, this theorem implies that no agent can be significantly influential at all, which contrasts sharply with the results in the previous literature. No matter how large the society is, an agent can no longer know the true state by observing a bounded number of actions (even if they are actions by close predecessors); however, an agent can and only can do so via observing an arbitrarily large neighborhood. Each agent makes a mistake with positive probability (when he decides not to observe), but efficient information aggregation occurs when the influence of any agent is arbitrarily small. The intuition behind this statement is that given any observed neighborhood, adding a different and sufficiently large neighborhood to it is always an \textit{ex ante} strict improvement on the posterior belief about the true state. Third, this result leads to a number of interesting comparative statics. For instance, in the limit, the equilibrium learning probability (the probability that an agent's action is correct) may be higher when the cost is positive than when the cost is zero and may be higher when private beliefs become weaker. Fourth, this theorem facilitates several important variations of the model. For instance, it can be shown that the pattern of social learning would be much different if endogenous observation preceded private signal, in contrast to the existing literature where this order of information makes no difference when observation is exogenous. In addition, a partial characterization of the level of social learning can be obtained under a more general cost structure.

The remainder of this paper is organized as follows: Section 2 provides a review of the related literature. Section 3 introduces the model. Section 4 defines the equilibrium and each type of learning that is discussed in this paper and characterizes the equilibrium behavior. Sections 5 to 7 present the main results and their implications, in addition to a number of extensions. Section 8 concludes. All the proofs are included in the Appendix.

\section{Literature Review}

A large and growing literature studies the problem of social learning by Bayesian agents who can observe others' choices. This literature begins with Bikhchandani, Hirshleifer and Welch\cite{BHW} and Banerjee\cite{Banerjee}, who first formalize the problem systematically and concisely and point to information cascades as the cause of herding behavior. In their models, the informativeness of the observed action history outweighs that of any private signal with a positive probability, and herding occurs as a result. Smith and Sorensen\cite{SS} propose a comprehensive model of a similar environment with a more general signal structure, and show that apart from the usual herding behavior, a new robust possibility of confounded learning occurs when agents have heterogeneous preferences: they neither learn the true state asymptotically nor herd on the same action. Smith and Sorensen\cite{SS} clearly distinguish ``private'' belief that is given by private signals and ``public'' belief that is given by observation, and they also introduce the concepts of bounded and unbounded private beliefs, whose meaning and importance were discussed above. These seminal papers, along with the general discussion by Bikhchandani, Hirshleifer and Welch\cite{BHW2}, assume that agents can observe the entire previous decision history, i.e., the whole ordered set of choices of their predecessors. This assumption can be regarded as an extreme case of exogenous network structure. In related contributions to the literature, such as Lee\cite{Lee}, Banerjee\cite{Banerjee2} and Celen and Kariv\cite{CK}, agents may not observe the entire decision history, but exogenously given observation remains a common assumption.

A more recent paper, Acemoglu et al.\cite{ADLO}, studies the environment where each agent receives a private signal about the underlying state of the world and observes (some of) their predecessors' actions according to a general stochastic network topology. Their main result states that when the private signal structure features unbounded belief, asymptotic learning occurs in each equilibrium if and only if the observation structure exhibits expanding observations. Other recent research in this area include Banerjee and Fudenberg\cite{BF}, Gale and Kariv\cite{GK}, Callander and Horner\cite{CH} and Smith and Sorensen\cite{SS2}, which differ from Acemoglu et al.\cite{ADLO} mainly in making alternative assumption for observation, i.e., that agents only observe the number of others taking each available action but not the positions of the observed agents in the decision sequence. However, all these papers also share the assumption of exogenous observation that is shared in the earlier literature discussed above.

The key difference between my paper and the literature discussed above is that observation is costly and strategic. First, each agent can choose whether to pay to acquire more information about the underlying state via observation. If the private signal is rather strong or the cost of observation is too high, an agent may rationally choose not to observe at all. Second, upon paying the cost, each agent can choose exactly which actions are included in the observation up to an exogenously given capacity constraint. In this way, society's observation network is endogenously formed, and hence we can examine not only the rational choice of action to match the true state but also the rational choice of whether to observe and which actions to observe as a cost-efficient decision regarding the acquisition of additional information. 

There have been several recent papers that discuss the impact of costly observation on social learning. In Kultti and Miettinen\cite{KM2}\cite{KM1}, both the underlying state and the private signal are binary, and an agent pay a cost for each action she observes. In Celen\cite{Celen}, the signal structure is similar to the general one adopted in this paper, but it is assumed that an agent can pay a cost to observe the entire action history before her. My model can be regarded as a richer treatment as compared to those papers, in the sense that it allows for a wide range of signal strucutures, as well as the possibility that agents would have to strategically choose a proper subset of their predecessors' actions to observe\footnote{For the main parts of this paper, agents are assumed to pay a single cost for observation, but in a later section I also present results regarding a more general cost function, which is non-decreasing in the number of actions observed.}. More importantly, this paper provides necessary and sufficient conditions for social learning to reach its theoretical upper bound under costly observation, which is a central question that remains unanswered in previous research. Some of the major findings in the above cited works, for example the existence of cost may lead to welfare improvement, are confirmed in this paper as well.

Another branch of the literature introduces a costly and strategic choice into the decision process -- each agent can pay to acquire an informative signal, or to ``search'', i.e., sample an available option and know its value. Notable works in this area include Hendricks, Sorensen and Wiseman\cite{HSW}, Mueller-Frank and Pai\cite{MP} and Ali\cite{Ali}. My paper differs from this stream of the literature in two aspects. On one hand, in those papers, the observation structure -- the neighborhood that each agent observes -- remains exogenous. In addition, agents in their models can obtain direct information about the true state such as signal or value of an option, whereas agents in the present paper can only acquire indirect information (others' actions) by paying the applicable cost.

There is also a well-known literature on non-Bayesian learning in social networks. In these models, rather than applying Bayes' update to obtain the posterior belief regarding the underlying state of the world by using all the available information, agents may adopt some intuitive rule of thumb to guide their choices (Ellison and Fudenberg\cite{EF}\cite{EF2}), only update their beliefs according to part of their information (Bala and Goyal\cite{BG}\cite{BG2}), or be subject to a certain bias in interpreting information (DeMarzo, Vayanos and Zwiebel\cite{DVZ}). Despite the various ways to model non-Bayesian learning, it is still common to assume that the network topology is exogenous. In terms of results, Golub and Jackson\cite{GJ} utilize a similar implication to that of Theorem 2 in this paper: they assume that agents naively update beliefs by taking weighted averages of their neighbors' beliefs and show that a necessary and sufficient condition for complete social learning (almost certain knowledge of the true state in a large and connected society over time) is that the weight put on each neighbor converges to zero for each agent as the size of the society increases.

Finally, the importance of observational learning via networks has been well documented in both empirical and experimental studies. Conley and Udry\cite{CU} and Munshi\cite{Munshi2} both focus on the adoption of new agricultural technology and not only support the importance of observational learning but also indicate that observation is often constrained because a farmer may not be able, in practice, to receive information regarding the choice of every other farmer in the area. Munshi\cite{Munshi} and Ioannides and Loury\cite{IL} demonstrate that social networks play an important role in individuals' information acquisition regarding employment. Cai, Chen and Fang\cite{CCF} conduct a natural field experiment to indicate the empirical significance of observational learning in which consumers obtain information about product quality from the purchasing decisions of others.

\section{Model}

\subsection{Private Signal Structure}

Consider a group of countably infinite agents: $\mathcal{N}=\{1,2,...\}$. Let $\theta\in\{0,1\}$ be the state of the world with equal prior probabilities, i.e., $Prob(\theta=0)=Prob(\theta=1)=\frac{1}{2}$. Given $\theta$, each agent observes an i.i.d. private signal $s_n\in S=(-1,1)$, where $S$ is the set of possible signals. The probability distributions regarding the signal conditional on the state are denoted as $F_0(s)$ and $F_1(s)$ (with continuous density functions $f_0(s)$ and $f_1(s)$). The pair of measures $(F_0,F_1)$ is referred to as the \textit{signal structure}, and I assume that the signal structure has the following properties:
\begin{itemize}
\item{1.} The pdfs $f_0(s)$ and $f_1(s)$ are continuous and non-zero everywhere on the support, which immediately implies that no signal is fully revealing regarding the underlying state.
\item{2.} Monotone likelihood ratio property (MLRP): $\frac{f_1(s)}{f_0(s)}$ is strictly increasing in $s$. This assumption is made without loss of generality: as long as no two signals generate the same likelihood ratio, the signals can always be re-aligned to form a structure that satisfies the MLRP.
\item{3.} Symmetry: $f_1(s)=f_0(-s)$ for any $s$. This assumption can be interpreted as indicating that the signal structure is unbiased. In other words, the distribution of an agent's private belief, which is determined by the likelihood ratio, would be symmetric between the two states.
\end{itemize}

Assumption 3 is strong compared with the other two assumptions. Nevertheless, many results in this paper can be easily generalized in an environment with an arbitrarily asymmetric signal structure. For those that do rely on symmetry, the requirement is not strict -- the results will hold as long as $f_1(s)$ and $f_0(-s)$ do not differ by very much, and an agent's equilibrium behavior is similar when receiving $s$ and $-s$ for a large proportion of private signals $s\in(-1,1)$. Therefore, the symmetry of signal structure serves as a simplification of a more general condition, whose essential elements are similar (in a symmetric sense) private signal distributions and similar equilibrium behavior under the two states. 

\subsection{The Sequential Decision Process}

The agents sequentially make a single action each between $0$ and $1$, where the order of agents is common knowledge. Let $a_n\in\{0,1\}$ denote agent $n$'s decision. The payoff of agent $n$ is
\begin{center}
$u_n(a_n,\theta)=\left\{
\begin{array}{c}
1,\text{ }if\text{ }a_n=\theta;\\
0,\text{ }otherwise.
\end{array}
\right.$
\end{center}

After receiving her private information and before engaging in the above action, an agent may acquire information about others from a network of observation\footnote{In Section 7.1, I will discuss the case involving an alternative order in which observation of actions takes place before a realization of private signal.}. In contrast with much of the literature on social learning, I assume that the network topology is not exogenously given but endogenously formed. Each agent $n$ can pay a cost $c\geq 0$ to obtain a \textit{capacity} $K(n)\in\mathbb{N}$; otherwise, he pays nothing and chooses $\varnothing$. I assume that the number of agents whose capacity is zero is finite, i.e., there exists $N\in\mathbb{N}$ such that $K(n)>0$ for all $n>N$.

With capacity $K(n)$, agent $n$ can select a \textit{neighborhood} $B(n)\subset\{1,2,\cdots,n-1\}$ of at most $K(n)$ size, i.e., $|B(n)|\leq K(n)$, and observe the action of each agent in $B(n)$. The actions in $B(n)$ are observed at the same time, and no agent can choose any additional observation based on what she has already observed. Let $\mathcal{B}(n)$ be the set of all possible neighborhoods of at most $K(n)$ size (including the empty neighborhood) for agent $n$. We say that there is a \textit{link} between agent $n$ and every agent in the neighborhood that $n$ observes. By the definition set forth above, a link in the network is \textit{directed}, i.e., unilaterally formed, and without affecting the observed agent. I refer to the set $\{K(n)\}_{n=1}^{\infty}$ as the capacity structure, and I define a useful property for it below.

\begin{defn}
A capacity structure $\{K(n)\}_{n=1}^{\infty}$ has \textbf{infinite observations} if
\begin{align*}
\lim_{n\rightarrow\infty}K(n)=\infty.
\end{align*}
If the capacity structure does not satisfy this property, then we say it has finite observations.
\end{defn}

\begin{exmp}
Some typical capacity structures are described below:
\begin{itemize}
\item{1.} $K(n)=n-1$ for all $n$: each agent can pay the cost to observe the entire previous decision history, which conforms to the early literature on social learning.
\item{2.} $K(n)=1$ for all $n$: each agent can observe only one of her predecessors. If observation is concentrated on one agent, the network becomes a star; at the other extreme, if each agent observes her immediate predecessor, the network becomes a line.
\end{itemize}
\end{exmp}

In between the above two extreme examples is the general case that $K(n)\in(1,n-1)$ for all $n$: each agent can, at a cost, observe an ordered sample of her choice among her predecessors. Note that a capacity structure featuring infinite observations requires only that the sample size grows without bound as the society becomes large but does not place any more restrictions on the sample construction. As will be shown in the subsequent analysis, this condition on sample size alone plays a key role in determining the achievable level of social learning.

An agent's strategy in the above sequential game consists of two problems: (1) given her private signal, whether to make costly observation and, if yes, whom to observe; (2) after observation (or not), which action to take between $0$ and $1$. Let $H_n(B(n))=\{a_m\in\{0,1\}:m\in B(n)\}$ denote the set of actions that $n$ can possibly observe from $B(n)$ and let $h_n(B(n))$ be a particular action sequence in $H_n(B(n))$. Let $I_n(B(n))=\{s_n,h_n(B(n))\}$ be $n$'s \textit{information set}, given $B(n)$. Agent $n$'s information set is her private information and cannot be observed by others\footnote{In Section 7.2, I will discuss the scenario with information diffusion, i.e., $h_n(B(n))$ can also be observed by creating a link with agent $n$.}. The set of all possible information sets of $n$ is denoted as $\mathcal{I}_n=\{I_n(B(n)):B(n)\subset\{1,2,\cdots,n-1\},|B(n)|\leq K(n)\}$.

A \textit{strategy} for $n$ is the set of two mappings $\sigma_n=(\sigma_n^1,\sigma_n^2)$, where $\sigma_n^1:S\rightarrow \mathcal{B}(n)$ selects $n$'s choice of observation for every possible private signal, and $\sigma_n^2:\mathcal{I}_n\rightarrow \{0,1\}$ selects a decision for every possible information set. A \textit{strategy profile} is a sequence of strategies $\sigma=\{\sigma_n\}_{n\in\mathbb{N}}$. I use $\sigma_{-n}=\{\sigma_1,\cdots,\sigma_{n-1},\sigma_{n+1},\cdots\}$ to denote the strategies of all agents other than $n$. Therefore, for any $n$, $\sigma=(\sigma_n,\sigma_{-n})$.

\subsection{Strong and Weak Private Beliefs}

An agent's \textit{private belief} given signal $s$ is defined as the conditional probability of the true state being $1$, i.e., $\frac{f_1(s)}{f_0(s)+f_1(s)}$. Note that it is a function of $s$ only, since it does not depend on the agents' actions. I now define several categories of private beliefs that will be useful in the subsequent analysis. The notions \textit{unbounded and bounded private beliefs} follow from the existing literature; the notions \textit{strong and weak private beliefs} are applicable for costly observation in particular.
\begin{itemize}
\item{1.} Agents have \textit{unbounded private beliefs} if
\begin{align*}
\lim_{s\rightarrow 1}\frac{f_1(s)}{f_0(s)+f_1(s)}&=1\\
\lim_{s\rightarrow -1}\frac{f_1(s)}{f_0(s)+f_1(s)}&=0.
\end{align*}
Agents have \textit{bounded private beliefs} if
\begin{align*}
\lim_{s\rightarrow 1}\frac{f_1(s)}{f_0(s)+f_1(s)}&<1\\
\lim_{s\rightarrow -1}\frac{f_1(s)}{f_0(s)+f_1(s)}&>0.
\end{align*}
The above definitions of unbounded and bounded private beliefs are standard in the previous literature and do not depend on the cost of observation $c$.
\item{2.} When $c>0$, agents have \textit{strong private beliefs} if there is $s^*<1$ and $s_*>-1$ such that
\begin{align*}
\frac{f_1(s^*)}{f_1(s^*)+f_0(s^*)}&=1-c\\
\frac{f_0(s_*)}{f_1(s_*)+f_0(s_*)}&=1-c.
\end{align*}
Given the symmetry assumption in the private signal structure, $s^*=-s_*$. Agents have \textit{weak private beliefs} if the above defined $s^*$ and $s_*$ do not exist.

Strong and weak private beliefs describe the relation between such private beliefs and the cost of observation. When an agent has strong private beliefs, she is not willing to pay the cost $c$ for a range of extreme private signals, even if doing so guarantees the knowledge of the true state of the world. In other words, private signals may have sufficiently high informativeness to render costly observation unnecessary. The opposite is weak private beliefs, in which case an agent always prefers observation when it contains enough information about the true state.
\end{itemize}

It is clear that unbounded private belief implies strong private belief for any positive cost. In the subsequent analysis, we will see that properties of private beliefs play an important role in the type of observational learning that can be achieved. To make the problem interesting, I assume that $c<\frac{1}{2}$; in other words, an agent will never choose not to observe merely because the cost is too high.

\section{Equilibrium and Learning}

\subsection{Perfect Bayesian Equilibrium}

Given a strategy profile, the sequence of decisions $\{a_n\}_{n\in\mathbb{N}}$ and the network topology (i.e., the sequence of the observed neighborhood) $\{B(n)\}_{n\in\mathbb{N}}$ are both stochastic processes. I denote the probability measure generated by these stochastic processes as $\mathcal{P}_\sigma$ and $\mathcal{Q}_{\sigma}$ correspondingly.

\begin{figure}[h]
\centering
\includegraphics[width=5.5in]{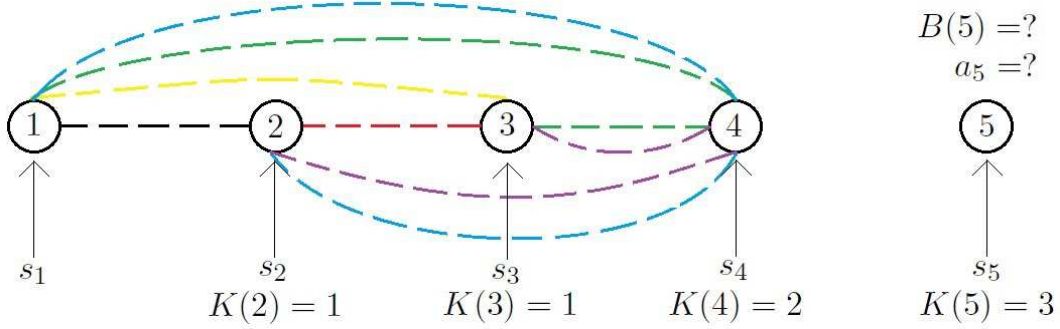}
\caption{Illustration of Network Topology}
\end{figure}

Figure 1 illustrates the world from the perspective of agent 5, who knows her private signal $s_5$, her capacity $K(5)$, and the possible observation behavior of predecessors $1-4$ (denoted by different colors). If agent $5$ knows the strategies of her predecessors, some possible behaviors may be excluded, e.g., agent $3$ may never observe agent $1$.

\begin{defn}
A strategy profile $\sigma^*$ is a pure strategy \textbf{perfect Bayesian equilibrium} if for each $n\in\mathbb{N}$, $\sigma_n^*$ is such that given $\sigma^*_{-n}$, (1) $\sigma_n^{*2}(I_n)$ maximizes the expected payoff of $n$ given every $I_n\in\mathcal{I}_n$; (2) $\sigma_n^{*1}(s_n)$ maximizes the expected payoff of $n$, given every $s_n$ and given $\sigma_n^{*2}$.
\end{defn}

For any equilibrium $\sigma^*$, agent $n$ first solves
\begin{align*}
\max_{y\in\{0,1\}}\mathcal{P}_{\sigma^*_{-n}}(y=\theta|s_n,h_n(B(n)))
\end{align*}
for any $s_n\in(-1,1)$ and any observed action sequence $h_n(B(n))$ from any $B(n)\subset\{1,\cdots,n-1\}$ satisfying $|B(n)|\leq K(n)$. This maximization problem has a solution for each agent $n$ because it is a binary choice problem. Denote the solution to this problem as $y_n^*(s_n,h_n(B(n)))$. Then, agent $n$ solves
\begin{align*}
\max_{B(n)\subset\{1,\cdots,n-1\}:|B(n)|\leq K(n)}\mathbb{E}[\mathcal{P}_{\sigma^*_{-n}}(y_n^*(s_n,h_n(B(n)))=\theta|s_n,h_n(B(n)))|s_n]
\end{align*}
This is, once again, a problem of discrete choice. Hence, given an indifference-breaking rule, there is a solution for every $s_n$. Finally, if the difference between the maximized expected probability of taking the correct action with observation and that without observation exceeds the cost of observation $c$, then the agent chooses to observe (the observed neighborhood being the solution to the above problem); otherwise, she chooses not to observe. Proceeding inductively for each agent determines an equilibrium.

Note that the existence of a perfect Bayesian equilibrium does not depend on the assumption of a symmetric signal structure. However, this assumption guarantees the existence of a \textit{symmetric} perfect Bayesian equilibrium, i.e., an equilibrium $\sigma^*$ in which, for each $s_n\in[0,1)$, $\sigma_n^{*1}(s_n)=\sigma_n^{*1}(-s_n)$. In other words, when the optimal neighborhood to observe (including the empty neighborhood, i.e., not to observe any predecessor) in equilibrium $\sigma^*$ is the same for every agent $n$, given any pair of private signals, $s_n$ and $-s_n$, then $\sigma^*$ is a symmetric equilibrium. In fact, if the optimal observed neighborhood is unique for each agent and every private signal, then each perfect Bayesian equilibrium will be symmetric. For instance, if for some private signal $s_4>0$, the unique optimal neighborhood for agent $4$ to observe is $\{2,3\}$, then due to the symmetric signal structure, it must also be her unique optimal neighborhood to observe when her private signal is $-s_4$. In the remainder of the paper, the analysis focuses on pure strategy symmetric Bayesian equilibria, and henceforth I simply refer to them as ``equilibria''. I note the existence of equilibrium below.

\begin{prop}
There exists a pure strategy symmetric perfect Bayesian equilibrium.
\end{prop}

As discussed briefly above, I will clearly identify which results can be generalized to an environment with asymmetric private signal distributions (and thus with asymmetric equilibria).

\subsection{Characterization of Individual Behavior}

My first results show that equilibrium individual decisions regarding whether to observe can be represented by an interval on the support of private signal.

\begin{prop}
When $c>0$, then in every equilibrium $\sigma^*$, for every $n\in\mathbb{N}$:
\begin{itemize}
\item{1.} For any $s_n^1>s_n^2\geq 0$ (or $s_n^1<s_n^2\leq 0$), if $\sigma_n^{*1}(s_n^1)\neq\varnothing$, then $\sigma_n^{*1}(s_n^2)\neq\varnothing$.
\item{2.} $\mathcal{P}_{\sigma^*}(a_n=\theta|s_n)$ is weakly increasing (weakly decreasing) in $s_n$ for all non-negative (non-positive) $s_n$ such that $\sigma_n^{*1}(s_n)\neq\varnothing$.
\item{3.} There is one and only one signal $s_*^n\in[0,1]$ such that $\sigma_n^{*1}(s_n)\neq\varnothing$ if $s_n\in[0,s_*^n)$ (if $s_n\in(-s_*^n,0]$) and $\sigma_n^{*1}(s_n)=\varnothing$ if $s_n>s_*^n$ (if $s_n<-s_*^n$).
\end{itemize}
\end{prop}

This proposition shows that observation is more favorable for an agent with a weaker signal, which is intuitive because information acquired from observation is relatively more important when an agent is less confident about her private information. The proposition then implies that for any agent $n$, there is one and only one non-negative cut-off signal in $[0,1]$, which is denoted as $s_*^n$, such that agent $n$ will choose to observe in equilibrium if $s_n\in[0,s_*^n)$ and not to observe if $s_n>s_*^n$. It is also clear that when $s_*^n\in(0,1)$, agent $n$ must be indifferent between observing and not observing at $s_n=s_*^n$. Under the symmetry assumption regarding the signal structure, the case when the private signal is non-positive is analogous. Figure 2 below illustrates the behavior of agent $n$ in equilibrium; note that when agent $n$ chooses to observe, the exact neighborhood observed may depend on her private signal $s_n$.

\begin{figure}[h]
\centering
\includegraphics[width=5.5in]{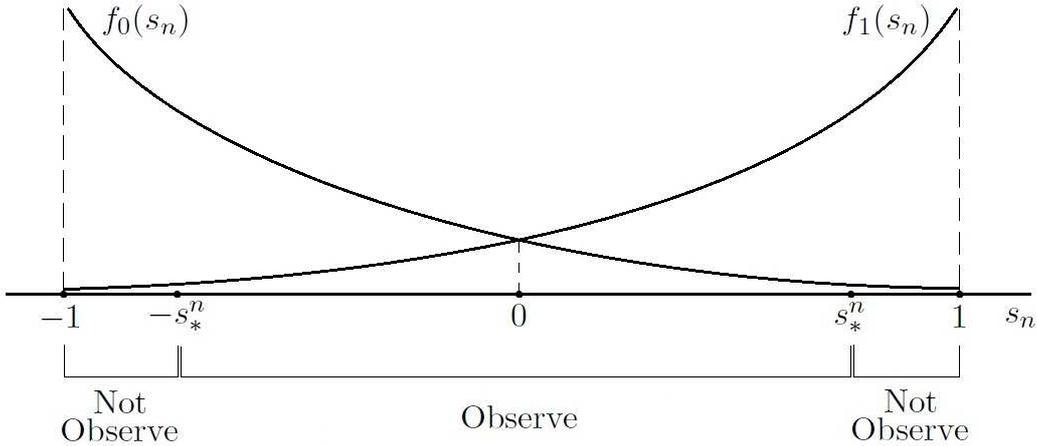}
\caption{Equilibrium Behavior of Agent $n$}
\end{figure}

The second implication of this proposition is that the learning probability (i.e., the probability of taking the correct action) has a nice property of monotonicity when the agent observes a non-empty neighborhood. When he chooses not to observe, i.e., when $s_n>s_*^n$ ($s_n<-s_*^n$), the probability of taking the correct action is also increasing (decreasing) in $s_n$ because the probability is simply equal to $\frac{f_1(s_n)}{f_0(s_n)+f_1(s_n)}$ ($\frac{f_0(s_n)}{f_0(s_n)+f_1(s_n)}$). However, this monotonicity is not preserved from observing to not observing because observation is costly and an agent with a stronger signal may be content with a lower learning probability to save on costs. Figure 3 below shows the shape of this probability with respect to $s_n$. Of the two continuous curves, the top curve depicts the probability of taking the correct action if agent $n$ always observes (denoted $P(a_n=\theta|O)$), whereas the bottom curve illustrates the probability of taking the correct action if agent $n$ never observes (denoted $P(a_n=\theta|NO)$). The solid ``broken'' curve measures the learning probability in equilibrium. The difference between the two continuous curves is greater than $c$ at $s_n\in[0,s_*^n)$ (and $s_n\in(-s_*^n,0]$), less than $c$ at $s_n>s_*^n$ (and $s_n<-s_*^n$), and equal to $c$ at $s_n=s_*^n$ (and $s_n=-s_*^n$).

\begin{figure}[h]
\centering
\includegraphics[width=5.5in]{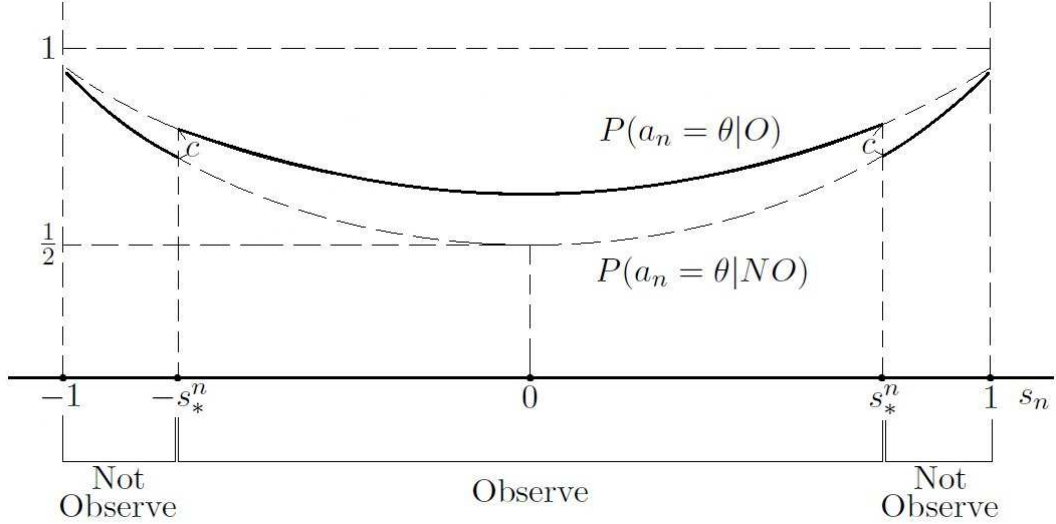}
\caption{Equilibrium Learning Probability for Agent $n$}
\end{figure}

\subsection{Learning}

The main focus of this paper is to determine what type of information aggregation will result from equilibrium behavior. First, I define the different types of learning studied in this paper.

\begin{defn}
Given a signal structure $(F_0,F_1)$, we say that \textbf{asymptotic learning} occurs in equilibrium $\sigma^*$ if $a_n$ converges to $\theta$ in probability: $\lim_{n\rightarrow\infty} \mathcal{P}_{\sigma^*}(a_n=\theta)=1$.
\end{defn}

Next, I define \textit{maximal learning}, which is a natural extension of asymptotic learning. Before the formal definition, I introduce an intermediate and conceptual term: suppose that a hypothetical agent can learn the true state by paying cost $c$. Clearly, an optimal strategy of this agent is to pay $c$ and learn the true state if and only if her private signal lies in some interval $(\underline{s},\bar{s})$ (this interval is equal to $(s_*,s^*)$ when private beliefs are strong and $(-1,1)$ when private beliefs are weak). Let $P^*(c)$ denote her probability of taking the right action under this strategy.

\begin{defn}
Given a signal structure $(F_0,F_1)$ and a cost of observation $c$, we say that \textbf{maximal learning} occurs in equilibrium $\sigma^*$ if the probability of $a_n$ being the correct action converges to $P^*(c)$: $\lim_{n\rightarrow\infty} \mathcal{P}_{\sigma^*}(a_n=\theta)=P^*(c)$.
\end{defn}

Asymptotic learning requires that the unconditional probability of taking the correct action converges to $1$, i.e., the posterior beliefs converge to a degenerate distribution on the true state. In terms of information aggregation, asymptotic learning can be interpreted as equivalent to making all private signals public and thus aggregating information efficiently. It marks the upper bound of social learning with an \textit{exogenous} observation structure. However, when observation becomes \textit{endogenous}, it is notable that asymptotic learning is impossible in certain cases. For instance, consider the case when cost is positive and private beliefs are strong. Indeed, because there is now a range of signals (to be precise, two intervals of extreme signals) such that an agent would not be willing to pay the cost even to know the true state with certainty, there is always the probability of making a mistake when the private signal falls into such a range. Therefore, an alternative notion is needed to characterize a more appropriate upper bound of social learning, which could theoretically be reached in some equilibrium. Maximal learning, as defined above, serves this purpose.

Maximal learning extends the notion of efficient information aggregation to an environment in which information acquisition is costly and means that, in the limit, agents can learn the true state as if they can pay the cost of observation to access all prior private signals, i.e., efficient information aggregation can be achieved at a price. From the perspective of equilibrium behavior, maximal learning occurring in an equilibrium implies that, in the limit, an agent will almost certainly take the right action whenever she chooses to observe. The term $P^*(c)$ is less than $1$ when private beliefs are strong\footnote{To be more precise, when private beliefs are strong, $P^*(c)=\frac{1}{2}F_0(s^*)+\frac{1}{2}(1-F_1(s_*))$. With a symmetric signal structure, it is equal to $F_0(s^*)$.} because an agent may choose not to observe when her private signal is highly informative. It is equal to $1$ when $c=0$ or when private beliefs are weak. In other words, maximal learning reduces to asymptotic learning in these two circumstances. The goal of this paper is then to characterize conditions that lead to maximal learning (or asymptotic learning, as a special case) in equilibrium.

\section{Learning with Zero Cost}

A central question is determining what conditions must be imposed on the capacity structure of observation for asymptotic/maximal learning. The answer to this is closely connected with the relation between the precision of private signals and the cost of observation. I begin by considering the case in which there is zero cost, i.e., observation is free. Even in this case, it is notable that not every agent will always choose to observe in equilibrium: if private signals are sufficiently strong, there may not be any realized action sequence in an observed neighborhood that can alter the agent's action. In other words, an agent may be indifferent between observation and no observation. 

The following theorem is one of the main results of the paper, and shows that unbounded private beliefs play a crucial role for learning in a society with no cost of observation. In particular, asymptotic learning, the strongest form of social learning, can be achieved in every equilibrium. This result holds even without the symmetry assumption for the signal structure.

\begin{thm}
When $c=0$ and agents have unbounded private beliefs, asymptotic learning occurs in every equilibrium.
\end{thm}

Theorem 1 presents an interesting comparison with the existing literature. Most of the above mentioned studies examine conditions on exogenous networks that induce (or do not induce) asymptotic learning; in contrast, Theorem 1 shows that as long as private beliefs are unbounded, a network topology that ensures asymptotic learning will \textit{automatically} form. In other words, with an endogenous network formation, the individual interest in maximizing the expected payoff and the social interest of inducing the agents' actions to converge to the true state are now aligned in the limit. In every equilibrium, agents with private signals that are not particularly strong will seek to increase the probability that they will take the right action via observation. Because there is no cost for observation, the range of signals given that an agent could choose to observe enlarges unboundedly within the signal support as the society grows. In the limit, each agent almost certainly chooses to observe, and information is thus efficiently aggregated without any particular assumption on the capacity structure.

The argument above also provides a general intuition behind the proof of this theorem. Suppose that asymptotic learning did not occur in some equilibrium, then there must be a limit to the probability of taking the correct action, whose value is less than one. Consider an agent $n$ whose learning probability is very close to this limit. For her immediate successor $n+1$, observing $n$'s action will produce a strict improvement: $n+1$ is as well-off as $n$ by following $n$'s action when $n+1$'s signal is not very precise, and strictly better-off than $n$ by following her own signal when the signal is very precise. Hence the limit on learning probability must be exceeded, a contradiction.

Acemoglu et al.\cite{ADLO} note that a necessary condition of network topology that leads to asymptotic learning is expanding observations, i.e., no agent is excessively influential in terms of being observed by others. In other words, no agent (or subset of agents) is the sole source of observational information for infinitely many other agents. This important result leads to the second implication of Theorem 1 regarding the equilibrium network topology: although it is difficult to precisely characterize agents' behavior in each equilibrium, we know that the equilibrium network must feature expanding observations, i.e., agents will always observe a close predecessor. This is an intuitive result because the action of someone later in the decision sequence presumably reveals more information. I formally describe this property of equilibrium network below.

\begin{cor}
If $c=0$ and agents have unbounded private beliefs, then every equilibrium $\sigma^*$ exhibits expanding observations: 
\begin{align*}
\lim_{n\rightarrow\infty}\mathcal{Q}_{\sigma^*}(\max_{b\in B(n)}b<M)=0\text{ for any }M\in\mathbb{N}.
\end{align*}
\end{cor}

A very simple but illustrative example of the foregoing is that of $K(n)=1$ for all $n$. As will be illustrated in details in the next section, the optimal observation of each agent (if any) in equilibrium must be the action of her immediate predecessor. The condition of expanding observations is satisfied exactly according to this description: because observation has no cost, each agent almost certainly chooses to observe in the limit, but no agent excessively influences other agents because each agent only influences her immediate successor. However, we will learn in the next section that when cost is positive and private beliefs are strong, an analogous condition -- observing a close predecessor when choosing to observe -- would \textit{not} suffice for the highest level of information aggregation achievable in equilibrium, i.e., for maximal learning.

In the other direction, when agents have bounded private beliefs, asymptotic learning does not occur for a number of typical capacity structures and associated equilibria. The following result lists some scenarios.

\begin{prop}
If $c=0$ and agents have bounded private beliefs, then asymptotic learning does not occur in the following scenarios:
\begin{itemize}
\item{(a)} $K(n)=n-1$.
\item{(b)} Some constant $\bar{K}$ exists such that $K(n)\leq \bar{K}$ for all $n$.
\end{itemize}
\end{prop}

The proposition above highlights two scenarios in which bounded private beliefs block asymptotic learning. In the first scenario, which corresponds to part $(a)$, it can be shown that the ``social belief'' for any agent, i.e., the posterior belief established from observation alone, is bounded away from $1$ in either state of the world, $0$ and $1$, regardless of the true state. As a result, asymptotic learning becomes impossible. With a positive probability, herding behavior occurs in equilibrium: either starting from some particular agent, \textit{all subsequent} agents choose the same (wrong) action (when social belief exceeds private belief at some point); or the equilibrium features longer and longer periods of uniform behavior, punctuated by increasingly rare switches (when social belief converges to but never exceeds private belief).

The second scenario, which corresponds to part $(b)$, posits that under either state, there is a positive probability that \textit{all} the agents choose incorrectly, which is another form of herding behavior. When private beliefs are bounded, an agent's private signal may not be strong enough to ``overturn'' the implication from a rather informative observation, and the agent would thus ignore her private information and simply follow her observation. This affects not only her own behavior but also the observational learning of her successors because they would also be aware that her action no longer reveals any information about her own private signal. Therefore, efficient information aggregation cannot proceed. For instance, it is clear that under either state of the world, the probability that the first $N$ agents all choose action $1$ is bounded away from zero. When $N$ is large, and when agent $N+1$ observes a large neighborhood such that an action sequence of $1$'s is more informative than each of her possible private signals, she will then also choose $1$ regardless of her own signal, and so will every agent after her. Herding behavior thus ensues as a result.

An atypical scenario that is not captured by the above two is that the limit superior of $K(n)$ is infinity but $K(n)<n-1$. It is still unclear whether asymptotic learning can occur in equilibrium in this case, because a full answer requires a complete characterization of the strategic selection of observed neighborhood by every agent. For the purpose of this paper, I refrain from discussing this topic in depth.

\section{Costly Learning with Strong Private Beliefs}

\subsection{Maximal Learning with Infinite Observations}

I have already argued before that when observation is costly and private beliefs are strong, asymptotic learning is impossible in any equilibrium. Furthermore, as will be shown below, maximal learning is not guaranteed in equilibrium either. In fact, we can see that whenever agents have finite observations, maximal learning cannot occur in any equilibrium. For an agent to choose to make a costly observation given her private signal, it must be the case that some realized action sequence in her observed neighborhood is so informative that she would rather turn against her signal and choose the other action. When private beliefs are strong, each agent chooses actions $0$ and $1$ with positive probabilities regardless of the true state; therefore, under either state $0$ or $1$, the above informative action sequence occurs with a positive probability. As a result, for any agent who chooses to observe, there is always a positive probability of making a mistake. For instance, consider again the example of $K(n)=1$ for all $n$. When private beliefs are strong, the probabilities that any agent would choose $0$ when $\theta=1$ and $1$ when $\theta=0$ are bounded below by $F_1(-s^*)$ and $1-F_0(s^*)$ correspondingly (with the symmetry assumption regarding the signal structure, the two probabilities are equal); therefore, the probabilities that any agent would take the wrong action when $\theta=1$ and when $\theta=0$ have the same lower bounds as well, given that this agent chooses to observe.

The next main result of this paper, Theorem 2, shows that a necessary and sufficient condition for maximal learning with strong private beliefs consists of infinite observations in the capacity structure.

\begin{thm}
When agents have strong private beliefs, maximal learning occurs in every equilibrium if and only if the capacity structure has infinite observations.
\end{thm}

The implication of Theorem 2 is two-fold. On one hand, the necessity of infinite observations stands in stark contrast to the expanding observations in the previous section, which means that no agent can be excessively influential but an agent may still be significantly influential for infinitely many others. In a world in which the cost of observation is positive and agents may sometimes rationally choose not to observe, for maximal learning to occur, \textit{no agent} can be significantly influential in the sense that any agent's action can only take up an arbitrarily small proportion in any other agent's observation. Indeed, because the probability of any agent making a mistake is now bounded away from zero, infinite observations must be required to suppress the probability of the wrong implication from an observed neighborhood.

On the other hand, Theorem 2 guarantees maximal learning when there are infinite observations. Whenever the size of the observed neighborhood can become arbitrarily large as the society grows, the probability of taking the right action based on observation converges to one. The individual choice of not observing, given some extreme signals -- and thus a source for any single agent to make a mistake on her own -- actually facilitates social learning by observation: because any agent may choose not to observe with positive probability, her action in turn must reveal some information about her private signal. Thus, by adding sufficiently many observations to a given neighborhood, i.e., by enlarging the neighborhood considerably, the informativeness of the entire observed action sequence can always be improved. Once a neighborhood can be arbitrarily large, information can be aggregated efficiently to reveal the true state.

Following this intuition, I now introduce an outline of the proof of Theorem 2 (detailed proofs can be found in the Appendix). Several preliminary lemmas are needed. The first lemma below simply formalizes the argument that when private beliefs are strong, each agent will choose not to observe with a probability bounded away from zero.

\begin{lem}
When agents have strong private beliefs, in every equilibrium $\sigma^*$, for all $n\in\mathbb{N}$, $s_*^n<s^*$.
\end{lem}

Next, I show that infinite observations are a necessary condition for maximal learning, in contrast to most existing literature stating that observing some close predecessor's action suffices for knowing the true state, in an exogenously given network of observation.

\begin{lem}
Assume that agents have strong private beliefs. If the capacity structure has finite observations, then maximal learning does not occur in any equilibrium.
\end{lem}

The logic behind the proof of Lemma 2 is rather straightforward. With strong private beliefs, in either state of the world, each agent takes actions $0$ and $1$ with probabilities bounded away from zero. Thus, when agent $n$ observes a neighborhood of at most $K$ size, the probability that the realized action sequence in this neighborhood would induce agent $n$ to take the wrong action is also bounded away from zero. If infinitely many agents can only observe a neighborhood whose size has the same upper bound, maximal learning can never occur.

The following few lemmas contribute to the proof of the sufficiency of infinite observations for maximal learning in every equilibrium. Given an equilibrium $\sigma^*$, let $B_k=\{1,\cdots,k\}$, and consider any agent who observes $B_k$. Let $R_{\sigma^*}^{B_k}$ be the random variable of the posterior belief on the true state being $1$, given each decision in $B_k$. For each realized belief $R_{\sigma^*}^{B_k}=r$, we say that a realized private signal $s$ and decision sequence $h$ in $B_k$ \textit{induce} $r$ if $\mathcal{P}_{\sigma^*}(\theta=1|h,s)=r$.

\begin{lem}
For either state $\theta\in\{0,1\}$ and for any $s\in(s_*,s^*)$, $\lim_{\epsilon\rightarrow 0^+}(\lim\sup_{k\rightarrow\infty}\mathcal{P}_{\sigma^*}(R_{\sigma^*}^{B_k}>1-\epsilon|0,s))=\lim_{\epsilon\rightarrow 0^+}(\lim\sup_{k\rightarrow\infty}\mathcal{P}_{\sigma^*}(R_{\sigma^*}^{B_k}<\epsilon|1,s))=0$.
\end{lem}

Lemma 3 shows that the action sequence in neighborhood $B_k$ cannot induce a degenerate belief on the wrong state of the world with positive probability as $k$ becomes large. This result is necessary because the posterior belief on the wrong state after observing the original neighborhood must be bounded away from $1$ if any strict improvement on the learning probability is to occur by expanding a neighborhood. In the next lemma, I demonstrate the feasibility of such strict improvement.

\begin{lem}
Assume that agents have strong private beliefs. Given any realized belief $r\in(0,1)$ on state $1$ for an agent observing $B_k$, for any $\hat{r}\in(0,r)$ ($\hat{r}\in(r,1)$), $N(r,\hat{r})\in\mathbb{N}$ exists such that a realized belief that is less than $\hat{r}$ (higher than $\hat{r}$) can be induced by additional $N(r,\hat{r})$ consecutive observations of action $0$ ($1$) in any equilibrium.
\end{lem}

Lemma 4 confirms the initial intuition that enlarging a neighborhood can strictly improve the informativeness of the observed action sequence. This improvement is represented by correcting a wrong decision by adding a sufficient number of observed actions. Moreover, the number of observed actions needed, $N(r,\hat{r})$, is independent of equilibrium. In the next lemma, I show that the strict improvement almost surely happens as $B_k$ becomes arbitrarily large, i.e., any posterior belief that leads to the wrong action will almost surely be reversed toward the true state after a sufficiently large number of actions are observed.

In the following lemma, given private signal $s$, let $\mathcal{P}_{\sigma^*}^{B_k}(\hat{a}\neq \theta|s)$ denote the probability of taking the wrong action for an agent observing $B_k$.

\begin{lem}
Assume that agents have strong private beliefs. Given any equilibrium $\sigma^*$ and any private signal $s\in(s_*,s^*)$, let $\hat{a}$ be the action that a rational agent would take after observing $s$ and every action in $B_k$. Then we have $\lim_{k\rightarrow\infty}\mathcal{P}_{\sigma^*}^{B_k}(\hat{a}\neq \theta|s)=0$.
\end{lem}

Lemma 5 is the most important lemma in the proof. It implies that a sub-optimal strategy -- observing the first $k$ agents in the decision sequence -- is already sufficient to reveal the true state when $k$ approaches infinity. Moreover, the sufficiency of this condition does not require any assumption regarding equilibrium strategies of the observed agents, which ensures its validity in every equilibrium. It then follows naturally that any agent's equilibrium strategy of choosing the observed neighborhood should generate a weakly higher posterior probability of taking the correct action. This argument is central for the proof of Theorem 2.

The key idea in Theorem 2 and its proof is that every observed action adds informativeness to the entire action sequence. The detailed proof shows that the symmetry assumption regarding the signal structure -- which leads to the existence of a symmetric equilibrium -- plays an important role in ensuring this condition. It has a natural interpretation: first, given that an agent makes a certain observation, because private signals are generated in an unbiased manner and agents behave similarly (in terms of choosing the observed neighborhood) when receiving symmetric signals, the Bayes's update by an observer of his action must be weakly in favor of the corresponding state as a result of the MLRP. Second, given that an agent chooses not to observe, the Bayes' update by the same observer would clearly strictly favor the corresponding state. These two effects combined show that the observation of every single action contributes a positive amount to information aggregation.

There are two special cases that ensure positive information contribution of each observed action, even with an asymmetric signal structure. The first case is when $s^*$ is sufficiently small or $s_*$ is sufficiently large. Intuitively, if an agent chooses not to observe given a relatively large range of private signals, her action should favor the corresponding state from a Bayesian observer's point of view, regardless of her behavior when she chooses to observe. In other words, the second effect mentioned above already suffices for a definitive Bayesian update, \textit{even without the symmetry assumption}. The following result formalizes this argument.

\begin{cor}
Given a general (potentially asymmetric) signal structure, if agents have strong private beliefs and $F_0(s_*)>F_1(s^*)$, then maximal learning occurs in every equilibrium if and only if the capacity structure has infinite observations.
\end{cor}

The second case is when there are infinitely many agents with observation capacity $K(n)=1$ or $K(n)=n-1$. In either of these two extreme cases, an agent's action reveals a significant amount of information about the true state, in the sense that the difference between the ex post belief before and after adding the observation of such an agent is always bounded away from zero. This property of information contribution only requires MLRP. Hence, once the capacity structure has infinite observations, observation consisting the actions of arbitrarily many these agents will result in maximal learning.

\begin{cor}
Given a general (potentially asymmetric) signal structure, if $\sharp\{n:K(n)=1\text{ or }K(n)=n-1\}=\infty$, then maximal learning occurs in every equilibrium if and only if the capacity structure has infinite observations.
\end{cor}

\subsection{An Example}

In this subsection, I introduce an example below to illustrate the difference among asymptotic learning, maximal learning and (a typical case of) learning in equilibrium with strong private beliefs and finite observations.

Assume the following signal structure:
\begin{align*}
F_0(s)&=\frac{1}{2}(s+1)(\frac{3}{2}-\frac{s}{2})\\
F_1(s)&=\frac{1}{2}(s+1)(\frac{1}{2}+\frac{s}{2}).
\end{align*}
This signal structure implies the probability density functions
\begin{align*}
f_0(s)&=\frac{1-s}{2}\\
f_1(s)&=\frac{1+s}{2}.
\end{align*}
Hence, it is easy to see that agents have unbounded (thus strong) private beliefs.

In addition, assume that $K(n)=1$ for all $n$. Consider the case when the cost of observation is low. When each agent can only observe one of her predecessors, if she chooses to observe then she would rationally choose to observe the agent with the highest probability of taking the right action. Agent 2 can only observe agent 1; agent 3, in view of this fact, will choose to observe agent 2 since agent 2's action is more informative than agent 1's action. Proceeding inductively, in every equilibrium, each agent will only observe their immediate predecessor when she chooses to observe, which results in a (probabilistic) ``line'' network. Let $\hat{s}^{*}=\lim_{n\rightarrow\infty} s_*^n$ and let $\hat{P}^*=\lim_{n\rightarrow\infty}\mathcal{P}_{\sigma^*}(a_n=\theta)$, it follows that the equations characterizing $\hat{s}^{*}$ and $\hat{P}^*$ are
\begin{align*}
&\hat{P}^*=F_0(-\hat{s}^*)+(F_0(\hat{s}^*)-F_0(-\hat{s}^*))\hat{P}^*\\
&\hat{P}^*-\frac{f_1(\hat{s}^{*})}{f_0(\hat{s}^{*})+f_1(\hat{s}^{*})}=c,
\end{align*}
The first condition decomposes the learning probability in the limit, $\hat{P}^*$, into the probability that an agent correctly follows his own signal without any observation, and the probability that she observes and her immediate predecessor's action is correct. The second condition indicates the indifference (in the limit) of an agent with signal $\hat{s}^*$ between observing and not observing her immediate predecessor, in the sense that the expected marginal benefit from observation must be equal to $c$. They can be further simplified as
\begin{align*}
\frac{1}{f_0(\hat{s}^{*})+f_1(\hat{s}^{*})}\frac{f_0(\hat{s}^{*})F_0(-\hat{s}^{*})-f_1(\hat{s}^{*})F_1(-\hat{s}^{*})}{F_0(-\hat{s}^{*})+F_1(-\hat{s}^{*})}=c.
\end{align*}
From the above equation and from the definition of strong private beliefs, we have
\begin{align*}
\hat{s}^{*}&=1-4c,\text{ if }c\leq\frac{1}{4}\\
s^*&=1-2c,\text{ if }c\leq\frac{1}{2}.
\end{align*}
It further implies that
\begin{align*}
\mathcal{P}_{\sigma^*}(a_n=\theta)&=1-4c^2,\text{ if }c\leq\frac{1}{4}\\
P^*(c)&=1-c^2,\text{ if }c\leq\frac{1}{2}.
\end{align*}
The first term $1-4c^2$ is the equilibrium probability of learning the true state in the limit when $K(n)=1$; the second term $1-c^2$ is the probability of learning the true state under maximal learning. Figure 4 illustrates the learning probability in the limit under asymptotic learning, under maximal learning and in equilibrium, as a function of the cost of observation $c$.

\begin{figure}[h]
\centering
\includegraphics[width=5.5in]{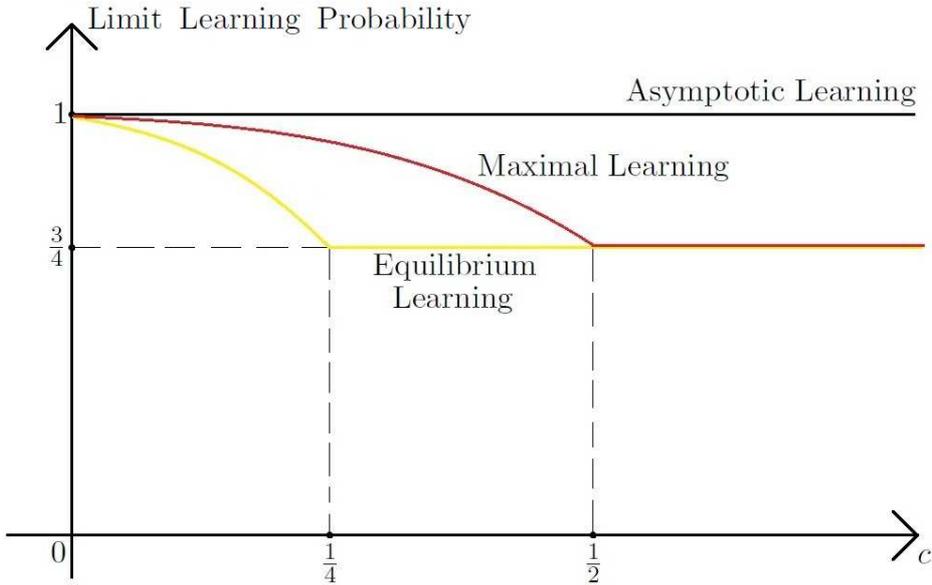}
\caption{Learning Probability as a Function of $c$}
\end{figure}

\subsection{Welfare Analysis}

In this subsection, I analyze the impact of observation cost and signal precision on the limit learning probability in equilibrium, $\lim_{n\rightarrow\infty}\mathcal{P}_{\sigma^*}(a_n=\theta)$. This probability represents the ultimate level of social learning achieved in a growing society. Two sets of parameters are of particular interest in this comparative statics: the cost of observation, $c$, and the precision of the private signal structure relative to cost. In many practical scenarios, these parameters capture the essential characteristics of a community with respect to how difficult it is to obtain information from others and how confident an agent can be about her private knowledge. The aim of this analysis is to identify the type(s) of environment that facilitate social learning.

In the following analysis, I show that compared to an environment with free observation, having a positive cost may actually improve the level of social learning. In previous sections, Theorem 1 shows that zero cost and unbounded private beliefs imply the highest learning probability, i.e., asymptotic learning; Theorem 2 allows us to obtain a straightforward formula for computing the limit learning probability with strong private beliefs and infinite observations:
\begin{align*}
\lim_{n\rightarrow\infty}\mathcal{P}_{\sigma^*}(a_n=\theta)=F_0(s^*)-F_0(-s^*)+F_0(-s^*)=F_0(s^*).
\end{align*}
Note that $F_0(s^*)-F_0(-s^*)$ is the probability (in the limit) that an agent chooses to observe; Theorem 2 indicates that observation reveals the true state of the world with near certainty when the society gets large. $F_0(-s^*)$ is the probability (in the limit) that an agent chooses not to observe and undertakes the correct action. My first result concerns an in-between case, i.e., under bounded private beliefs and infinite observations, the comparison between an environment with zero cost and one with positive cost. For any single agent, other things equal, it is always beneficial to observe with no cost than with positive cost. However, positive cost may actually be desirable for the society as a whole: for any agent, even though relying on her signal more often is harmful to her own learning, it provides more information for her successors who observe her action, hence raising the informativeness of observation. This argument provides the intuition for the formal result below.

Consider the capacity structure $K(n)=n-1$, i.e., any agent is able to observe all her predecessors. Let $\sigma^*(c)$ be an equilibrium under cost $c$, and let $P^*(\sigma^*(c))$ be the limit probability of learning, given $\sigma^*(c)$, i.e., $P^*(\sigma^*(c))=\lim_{n\rightarrow\infty}\mathcal{P}_{\sigma^*(c)}(a_n=\theta)$.

\begin{prop}
Assume that agents have bounded private beliefs. Let $\sigma^*(0)$ be any equilibrium under zero cost. There are positive values $\bar{c},\underline{c}$ ($\bar{c}>\underline{c}$), such that for any $c\in(\underline{c},\bar{c})$ and any $\sigma^*(c)$, $P^*(\sigma^*(0))<P^*(\sigma^*(c))$.
\end{prop}

With zero cost and bounded private beliefs, herding occurs because at some point in the decision sequence, an agent may abandon \textit{all} her private information, although her observation is not perfectly informative of the true state. Yielding to observation, in turn, causes her own actions to reveal no information about her private signal, and thus information aggregation ends. However, with positive cost and strong private beliefs -- and although nothing has changed in the signal structure -- now every agent relies on \textit{some} of her possible private signals, which strengthens the informativeness of observation. When the probability of an agent choosing to observe is sufficiently high (but still bounded away from one), an agent may enjoy a higher chance of taking the right action than when observation is free for everyone. The comparison is illustrated in Figure 5.

\begin{figure}[h]
\centering
\includegraphics[width=5.5in]{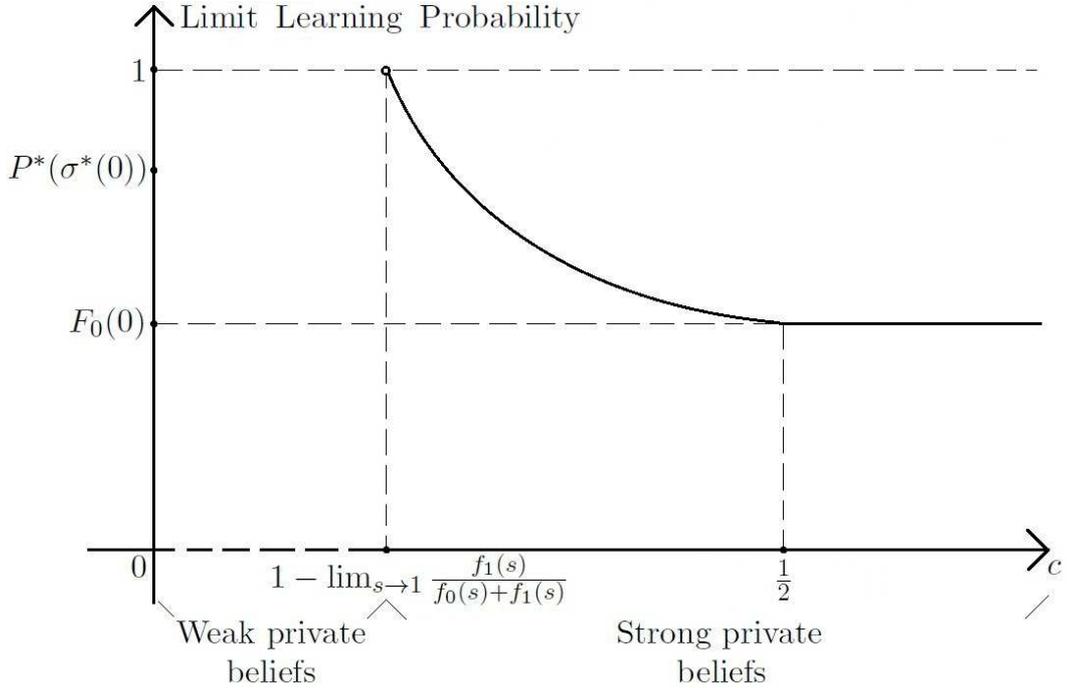}
\caption{Limit Learning Probability and Cost of Observation}
\end{figure}

Next, I consider the effect of signal strength that is measured by the probability of receiving relatively more informative private signals. In most existing literature, the network of observation is exogenously given. In other words, observation is ``free'' and non-strategic, which is not affected by how accurate an agent's private signal is. However, when observation becomes strategic and costly, there is a trade-off between obtaining a higher probability of taking the right action and saving the cost. As a result, when an agent receives a rather strong signal, she might as well cede the opportunity of observational learning and just act according to her private information. Therefore, in this environment, strong signals can be detrimental to social learning. The next result demonstrates this phenomenon. 

With strong private beliefs, denote the \textit{strength} of the private signal relative to cost as $F_0(-s^*)+1-F_0(s^*)$, i.e., the probability of not observing even if observing reveals the true state.

\begin{prop}
Generically, there exist scenarios where a signal structure with higher strength leads to lower limit learning probability.
\end{prop}

In summary, we only see clear monotonicity (lower cost or stronger signals are better for social learning) in extreme scenarios (unbounded private beliefs or zero cost). When private beliefs are bounded and cost is positive, two new factors enter the determinant of the limit learning probability. First, the fact that costly observation alone may now provide higher informativeness than free observation implies that positive cost may actually be more favorable for social learning. Second, positive cost signifies a trade-off between two components of an agent's final payoff -- the probability of undertaking the right action and the cost of observation -- thus, from the perspective of social learning, weaker private signals may be preferred because they incentivize agents to achieve better learning by observation. As a result of these joint effects, the learning process via endogenous networks of observation becomes more complex.

\subsection{Flexible Observations with Non-Negative Marginal Cost}

Thus far, I have assumed a single and fixed cost for observing any neighborhood of size up to the capacity constraint. Another interesting setting is to assume that the cost of observation depends on the number of actions observed. It can be easily anticipated that a full characterization of the pattern of social learning is difficult, given an arbitrary cost function; such characterization requires detailed calculations regarding the marginal benefit of any additional observed action, which varies substantially based on the specific signal distributions. However, in the following typical class of cost functions, the previous results can easily be applied to describe the level of social learning when the cost of observation changes with the number of observed actions.

Consider the following setting: after receiving her private signal, each agent can decide how many actions (up to $K(n)$) to observe. As in the original model, the actions are observed simultaneously\footnote{Nevertheless, the results below still hold in the context of sequential observation, i.e., an agent can choose whether to observe an additional action by paying the marginal cost, based on what she has already observed.}. The cost function of observing $m$ actions is denoted with $c(m)$. Assume that $c(0)=0$ and that $c(m)$ satisfies the property of non-negative marginal cost: $c(m+1)-c(m)\geq 0$ for all $m\in\mathbb{N}$. Here, maximal learning is defined as to achieve efficient information aggregation in the limit by paying the least cost possible: $\lim_{n\rightarrow\infty}\mathcal{P}_{\sigma^*}(a_n=\theta)=P^*(c(1))$. The following result is essentially a corollary of Theorems 1 and 2 and characterizes the pattern of social learning under this class of cost functions.

\begin{prop}
Assume that agents have unbounded private beliefs. Under the above class of cost functions, the following propositions are true for the social learning process:
\begin{itemize}
\item{(a)} Asymptotic learning occurs in every equilibrium if and only if $c(1)=0$.
\item{(b)} When $c(1)>0$, maximal learning occurs in every equilibrium if $c(m+1)-c(m)=0$ for all $m\geq 1$; otherwise, maximal learning does not occur in any equilibrium.
\end{itemize}
\end{prop}

Proposition 6 indicates that the difference in the level of social learning produced by the different costs of observation becomes even larger in the context of a more general cost function. First, the results show that the key factor determining whether asymptotic learning occurs in equilibrium is $c(1)$, i.e., the cost of observing the first action. When $c(1)=0$, any agent can at least do as well as any of her predecessors by simply observing the latter and following the observed action; thus, when private beliefs are unbounded, learning does not stop until agents almost certainly undertake the correct action. Secondly, when $c(1)>0$, maximal learning occurs if and only if each additional observation is free. We already know that maximal learning requires agents to choose to observe arbitrarily large neighborhoods when the society is large; with non-negative marginal cost, the cost of observing such a neighborhood gets strictly higher than $c(1)$ as long as the marginal cost of observing some other action is positive. As a result, even if the capacity structure has infinite observations, an agent will choose not to observe when she receives a signal that makes her more or less indifferent between not observing and paying $c(1)$ to know the true state. Therefore, maximal learning never occurs.

\section{Discussion}

\subsection{Observation Preceding Signal}

In the previous analysis, we see that under strong private beliefs, (1) asymptotic learning is impossible and (2) if the capacity structure has only finite observations and observation is costly, then maximal learning does not occur in any equilibrium even when private beliefs are unbounded. As it turns out, an agent's \textit{timing} of choosing her observation plays an important role: because an agent receives her private signal before choosing observations, it always remains possible that an agent chooses not to observe and bases her action solely on her private signal, which may be rather strong but is nonetheless not perfectly informative. Thus, when observations are finite, there is always a probability bounded away from zero that observations will induce incorrect action.

However, in practical situations, the timing of the arrival of different types of information is often not fixed. For instance, when a firm decides whether to adopt a new production technology, it may well take less time to conduct a survey about which nearby firms have already implemented the technology than to obtain private knowledge about the technology itself via research and trials. It is then interesting to study the different patterns that the social learning process would exhibit under this alternative timing. The next result demonstrates that when observation precedes private signal, learning somehow becomes easier as long as the cost of observation is not too high: asymptotic learning can occur even when cost is positive and observations are finite. Such difference between timing schemes only arises when observation is endogenous and costly -- when observation is exogenous or free, the two timing schemes would essentially lead to identiical equilibria.

Consider the alternative dynamic process in which each agent chooses her observed neighborhood before receiving her private signal. Let $Y(m)$ denote the probability that an agent will take the right action if she can observe a total of $m$ other agents, each of whose actions are based solely on her own private signal. The result can be easily generalized to the case with an asymmetric signal structure.

\begin{prop}
When agents have unbounded private beliefs, asymptotic learning occurs in every equilibrium if and only if there exists $n$ such that $Y(K(n))-F_0(0)\geq c$.
\end{prop}

Proposition 7 shows that a necessary and sufficient condition for asymptotic learning is the existence of an agent who initiates the information aggregation process by beginning to observe. Because observation precedes the private signal, each of her successors can be at least as well off as she is simply by observing her action. Therefore, after this starting point, observation becomes the optimal choice even for agents with lower capacity. Furthermore, because there is no conflict between a strong signal and costly observation (observation occurs first anyway), there is no blockade of information once observation begins. As in the case with unbounded private beliefs and zero cost, a network topology featuring expanding observations will spontaneously form and asymptotic learning will occur as a result.

For better illustration, consider an environment with unbounded private beliefs and infinite observations in which the limit learning probability can be fully characterized both when a signal precedes observation and when observation precedes a signal. Figure 6 shows the relation between the limit learning probability and the cost of observation under the two timing schemes; this figure also shows that asymptotic learning occurs within a much larger range of cost when observation comes first, while the limit learning probability falls abruptly to that with no observation when cost becomes high because of the lack of an agent to trigger observational learning. When agents receive private signals first, the limit learning probability is continuous against the cost of observation, and the threshold of cost above which observation stops is higher; as a result, agents learn less when cost is relatively low and more when cost is intermediate compared with the other timing scheme.

\begin{figure}[h]
\centering
\includegraphics[width=5.5in]{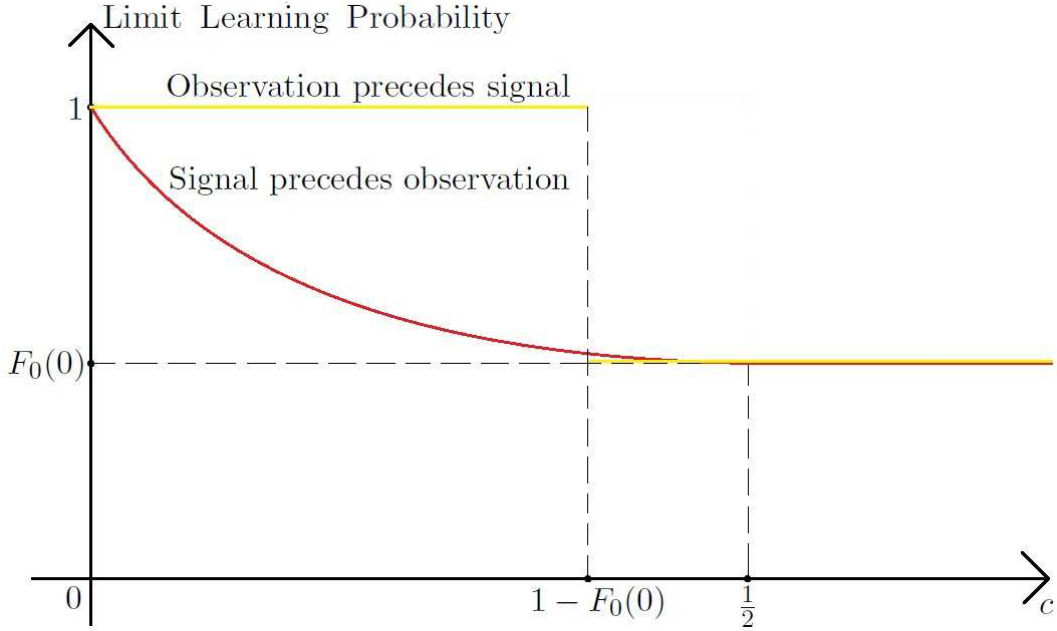}
\caption{Limit Learning Probability under Two Timing Schemes}
\end{figure}

When private beliefs are bounded, a partial characterization analogous to Proposition 3 can be obtained under this alternative timing: asymptotic learning fails for a number of typical capacity structures and associated equilibria. In each scenario, as with the analysis above, once observation is initiated by some agent, all the successors will choose to observe. Of course, the cost of observation must be bounded by a certain positive value (which can be characterized based on the specified equilibrium behavior) to ensure the existence of the particular equilibrium; otherwise, observation never begins and each agent would simply act in isolation according only to her private signal.

\subsection{Information Diffusion}

Another important assumption in the renowned herding behavior and information cascades literature is that observing an agent's action does not reveal any additional information regarding the agent's knowledge about the actions of others, which occurs without much loss of generality in earlier models because agents are assumed to observe the entire past action history in any event. In a more general setting, it can be expected that allowing an agent to access the knowledge (still about actions and not about private signals) of agents in their observed neighborhood makes a significant difference because information can now flow not only through direct links in the network but also through indirect paths. In this section, I discuss the impact of such added informational richness on the level of social learning.

Assume the following \textit{information diffusion} in the observation structure: if agent $n$ has observed the actions in neighborhood $B(n)$ before choosing her own action, then any agent observing $n$ knows $a_n$ and each action in $B(n)$. In our model of endogenous network formation, this alternative assumption has a particular implication: if agent $n$ sees that another agent $m$ chose action $1$ but $m$ did not know the action of anyone else, $n$ can immediately infer that $m$ must have received a rather strong signal. As a result, when the observed neighborhood becomes large, the observing agent can apply the weak law of large numbers to draw inferences regarding the true state of the world. With this simpler argument, the symmetry assumption regarding the signal structure can be relaxed to obtain the following result.

\begin{prop}
With information diffusion, when agents have strong private beliefs, maximal learning occurs in every equilibrium if and only if the capacity structure has infinite observations.
\end{prop}

Somewhat curiously, introducing information diffusion into the model only leads to relaxing the symmetry assumption but still results in the same necessary and sufficient condition for maximal learning. The underlying reason for this result is that as long as each agent chooses only to observe with a probability bounded away from 1 (i.e., for a range of signals that differs significantly from full support), every agent will only know the finite actions of others with near certainty if observations are finite. Thus, when the capacity structure has finite observations, the additional information acquired via information diffusion will not be sufficient for maximal learning.

We have seen from the above result that even with information diffusion, a capacity structure with finite observations never leads to maximal learning in any equilibrium. An even more surprising observation is that, when the capacity structure has finite observations, information diffusion may not be helpful at all in terms of the limit learning probability. For instance, consider the capacity structure $K(n)=1$ for all $n$ and symmetric private signals, and consider agent $n$ where $n$ is large. In equilibrium, if any agent chooses to observe, she will observe her immediate predecessor. By choosing to observe some agent $m_1$, agent $n$ will know the actions of an almost certainly finite ``chain'' of agents $m_1,m_2,\cdots,m_l$, such that $m_1$ observed $m_2$, $\cdots$, $m_{l-1}$ observed $m_l$, and $m_l$ chose not to observe. It is first clear that $a_{m_{l-1}}$ must almost surely equal $a_{m_l}$ because the range of signals -- given that $m_{l-1}$ chooses to observe -- is close to that for $m_l$ (by the assumption that $n$ is large), which induces $m_{l-1}$ to follow $m_l$'s action that implies a stronger private signal. Hence, $m_{l-1}$'s action does not reveal any additional information about the true state. Repeating this argument inductively, the only action that is informative to agent $n$ is $a_{m_l}$. If agent $n$ can only observe a single action, she can use the identical argument to deduce that the observation ultimately reflects the action of an agent who chose not to observe. Therefore, the limit learning probability $\lim_{n\rightarrow\infty}\mathcal{P}_{\sigma^*}(a_n=\theta)$ is not affected by information diffusion.

\subsection{Costly Learning with Weak Private Beliefs}

When private beliefs are weak, asymptotic/maximal learning may not be achieved in the previous model because of the possibility of herding. More specifically, as $n$ becomes large, some agents' actions may be so accurate (although not perfect) that successors choose to observe and follow them regardless of their own private signal. Hence, more observations do not necessarily provide more information about the true state when the observation structure is \textit{determinant}, i.e., any agent $n$ can choose their neighborhood among $\{1,\cdots,n-1\}$. A partial characterization similar to Proposition 3 can be obtained that shows that maximal learning (which is equivalent to asymptotic learning when private beliefs are weak) cannot occur for a number of typical capacity structures and the associated equilibria. However, maximal learning can still be approximated with certain general \textit{stochastic} observation structures. First, I introduce the notion of $\epsilon$-maximal learning, which is the approximation of the original notion of maximal learning. 

\begin{defn}
Given a signal structure $(F_0,F_1)$, we say that \textbf{$\epsilon$-maximal learning} occurs in equilibrium $\sigma^*$ when the limit inferior of the probability of $a_n$ being the correct action is at least $(1-\epsilon)P^*(c)$: $\lim\inf_{n\rightarrow\infty} \mathcal{P}_{\sigma^*}(a_n=\theta)\geq (1-\epsilon)P^*(c)$.
\end{defn}

$\epsilon$-maximal learning describes the situation in which an agent undertakes the correct action with a probability of at least $(1-\epsilon)P^*(c)$. Note that $P^*(c)=1$ when private beliefs are weak. Hence, $\epsilon$-maximal learning under weak private beliefs implies that the learning probability $\mathcal{P}_{\sigma^*}(a_n=\theta)$ is bounded below by $1-\epsilon$ in the limit. In fact, when $\epsilon$ is close to zero, social learning will be almost asymptotic. I will provide sufficient conditions below at an equilibrium for $\epsilon$-maximal learning to occur. The essential factor that facilitates $\epsilon$-maximal learning is the existence of a \textit{non-persuasive neighborhood}, which is introduced by Acemoglu et al.\cite{ADLO}. I define this concept below.

\begin{defn}
When agents have weak private beliefs, let
\begin{align*}
\bar{\beta}&=\lim_{s\rightarrow +1}\frac{f_1(s)}{f_0(s)+f_1(s)}\\
\underline{\beta}&=\lim_{s\rightarrow -1}\frac{f_0(s)}{f_0(s)+f_1(s)}
\end{align*}
denote the upper and lower bounds of an agent's private beliefs on the true state being $1$. A finite subset of agents $B$ is a \textbf{non-persuasive neighborhood} in equilibrium $\sigma^*$ if
\begin{align*}
\mathcal{P}_{\sigma^*}(\theta=1|a_k=y_k\text{ for all }k\in B)\in (\underline{\beta},\bar{\beta})
\end{align*}
for any set of values $y_k\in\{0,1\}$ for each $k$.
\end{defn}

A neighborhood $B$ is non-persuasive with respect to equilibrium $\sigma^*$ if given any possible realized action sequence in this neighborhood, an agent that observes it may still rely on his own private signal. If a neighborhood is non-persuasive, then any agent will choose not to observe it and follow his own private signal with positive probability. In other words, regardless of the realized action sequence, there exist a positive measure of private signals such that the agent takes action $0$ and another positive measure of private signals such that the agent takes action $1$. Note that the definition of a non-persuasive neighborhood depends on the particular equilibrium $\sigma^*$; however, there is a class of neighborhoods that are non-persuasive in any equilibrium. Suppose that $K(n)=0$ for $n=1,2,\cdots,M$ for some $M\in\mathbb{N}^+$; in other words, the first $M$ agents cannot observe any of the actions of others. In this case, there is a positive $M'\leq M$ such that any subset of $\{1,\cdots,M'\}$ is a non-persuasive neighborhood in any equilibrium. To illustrate this, simply note that $\{1\}$ already satisfies the criterion: for any agent who can only observe agent $1$'s action, there must be a range of strong private signals that are more informative about the true state than the action of agent $1$.

Consider the following stochastic observation structure for equilibrium $\sigma^*$: there are $M$ non-persuasive neighborhoods $C_1,\cdots,C_M$ such that for all $n$, agent $n$ can only observe within some $C_i$, $i\in\{1,\cdots,M\}$, with probability $\epsilon_n>0$; with probability $1-\epsilon_n$, agent $n$ can observe within $\{1,\cdots,n-1\}$. In both cases, the capacity structure $\{K(n)\}_{n=1}^{\infty}$ stays the same. Proposition 9 below provides a class of stochastic observation structures in which $\epsilon$-maximal learning occurs for any signal structure. Establishing this result depends on how to interpret an observed action from a Bayesian observer's perspective. Given the positive probability with which an agent can only choose to observe a non-persuasive neighborhood, her final action now may reflect either conformity with her observation or her strong private signal. This property holds even when private beliefs are weak. Hence, we can follow a similar argument to that in the proof of Theorem 2 that shows that the total informativeness of the action sequence in any neighborhood can always be increased by including in it enough additional actions.

\begin{prop}
In any equilibrium $\sigma^*$ with the above stochastic capacity structure, $\epsilon$-maximal learning occurs if $\lim_{n\rightarrow\infty}\epsilon_n=\epsilon$ and $\lim_{n\rightarrow\infty}K(n)=\infty$.
\end{prop}

The implication of Proposition 9 is rather surprising. In contrast to many existing results in the literature that lead to herding behavior more easily when agents receive weaker signals, this theorem indicates that learning dynamics can be much richer and social learning can be rather accurate in a general network of observation in which links are formed endogenously. In particular, the limit learning probability can even be higher under weak private beliefs than under strong private beliefs. For example, in an environment with infinite observations, consider an equilibrium in which beginning with some agent $n$, each agent chooses to observe regardless of her private signal when she can observe within the whole set of her predecessors (such an equilibrium can exist only under weak private beliefs). Proposition 9 shows that the learning probability $\mathcal{P}_{\sigma^*}(a_n=\theta)$ is bounded below by $1-\epsilon$ in the limit because the probability that an agent chooses to observe has the same lower bound in the limit. When $\epsilon$ is arbitrarily small, the agents can almost asymptotically learn the true state of the world, which they cannot do under strong private beliefs and which also relates to the result that weaker private signals may actually imply a higher limit learning probability in the welfare analysis of the previous section.

Comparing Proposition 9 with existing results in the literature on achieving asymptotic learning under bounded private beliefs in an exogenously given and stochastic observation structure (e.g., Theorem 4 in Acemoglu et al.\cite{ADLO}) is also illuminating. The main implication from existing results is that, if agents observe the entire history of actions with some probability (which is uniformly bounded away from zero) and observe some non-persuasive neighborhood with some probability (which converges to zero but when the infinite sum of such probabilities over agents is unbounded) in a stochastic observation structure satisfying expanding observations, then observing some close predecessor reveals the true state in the limit. In Proposition 9, the role of observing a non-persuasive neighborhood with positive probability is similar -- to make agents rely on their own signals with positive probability such that their actions become informative -- but to know the true state in the limit, the key factor continues to be to observe an arbitrarily large neighborhood. Thus, to approximate maximal learning with endogenous observation, even when $\lim_{n\rightarrow\infty}K(n)=\infty$, it is not sufficient for the probability of making observations within $\{1,\cdots,n-1\}$ to be bounded away from zero; instead, such probability must be close to one. 

As a useful sidenote, the results in this paper also apply in the generalized context in which agents are divided into groups $g_1,g_2,\cdots$. In period $i\in\mathbb{N}$, agents in group $g_i$ make their choices simultaneously with capacity $K(i)$. In other words, the largest neighborhood that an agent in $g_i$ can observe is $\cup_{j=1}^{i-1}g_j$. This generalization allows for the possibility that multiple agents move (choose their observed neighborhood and then their action) simultaneously during each period.

\section{Conclusion}

In this paper, I have studied the problem of sequential learning in a network of observation. A large and growing literature has studied the problem of social learning in exogenously given networks. Seminal papers, such as Bikhchandani, Hirshleifer and Welch\cite{BHW}, Banerjee\cite{Banerjee} and Smith and Sorensen\cite{SS}, first studied environments in which each agent can observe the entire past history of actions. Recent contributions, such as Acemoglu et al.\cite{ADLO}, generalized the network topology to being stochastic. The central question is whether equilibria lead to asymptotic learning, i.e., efficient aggregation of information, and the results in the literature point to two crucial factors: unbounded private beliefs, and expanding observations (i.e. no agent is excessively influential). When these two criteria are satisfied, asymptotic learning occurs in every equilibrium.

In many relevant situations, individuals do not automatically acquire information from a given network of observation but are instead engaged in strategic and costly observations of others' actions. Such behavior can be understood as forming links with others, which constitutes the ultimate network of observation. This raises the question of how information is aggregated in an endogenously formed network and what level of social learning can be achieved in equilibrium in different scenarios. To address these questions, I have formulated a model of sequential learning in a network of observation constructed by agents' strategic and costly choices of observed neighborhoods.

In the model, agents sequentially make strategic moves. Each agent receives an informative private signal, after which she can pay a cost to observe a neighborhood among her predecessors. The size of the observed neighborhood is limited by a given capacity constraint. Given her signal and the realized action sequence in her chosen neighborhood, each agent then chooses one of two possible actions. I have characterized pure-strategy perfect Bayesian equilibria for arbitrary capacity structures and have also characterized the conditions under which different types of social learning occur. In particular, I focused on asymptotic learning and maximal learning. Asymptotic learning refers to efficient information aggregation, i.e. agents' actions converging in probability to the right action as the society becomes large. Maximal learning refers to the same convergence that is conditional on observation. Maximal learning reduces to asymptotic learning when agents almost certainly choose to observe.

Two concepts are shown to be crucial in determining the level of social learning in equilibrium. The first is the precision of private signals. Apart from whether private beliefs are bounded, the relation between private beliefs and the cost of observation is equally important. When the cost is positive, I say that private beliefs are strong if for some signals an agent would not be willing to pay the cost even to know the true state, and that otherwise private beliefs are weak. The second important concept is that of infinite or finite observations with respect to the capacity structure.

My first main result, Theorem 1, shows that when private beliefs are unbounded and the cost of observation is zero, asymptotic learning occurs in every equilibrium. It further implies that the network topology in every equilibrium will automatically feature expanding observations. It provides a micro-foundation for the above mentioned condition of expanding observation, i.e. agents tend to observe some close predecessor when observation is not costly. In this case, information is always efficiently aggregated unconditionally.

The next main theorem, Theorem 2, characterizes the necessary and sufficient condition for maximal learning when agents have strong (not necessarily unbounded) private beliefs: if and only if the capacity structure has infinite observations, agents will learn the true state with near certainty via observation when the society becomes large. This result stands in stark contrast to the literature in the sense that each agent must be infinitesimally influential to others in any observed neighborhood to ensure maximal learning. As long as there are ``influential'' agents, even though they may not be excessively influential, there is always a positive probability of taking the wrong action after observation. Conversely, whenever observations are infinite, information can be aggregated efficiently to guarantee the revelation of the true state conditional on observation.

I believe that the framework developed in this paper has the potential to facilitate a more general analysis of sequential learning dynamics in an endogenously formed network of observation. The following questions are among those that can be studied in future work using this framework: (1) equilibrium learning when agents' preferences are heterogeneous; (2) equilibrium learning when the cost of observation is random and is part of an agent's private information; (3) the relation between the cost of observation and the speed (rate of convergence) of sequential learning.

\newpage
\appendix
\section*{APPENDIX}

\begin{proof}[Proof of Proposition 2]
1: Consider any $s_n\geq 0$. Let $H_n^1(s_n)$ ($H_n^0(s_n)$) denote the set of observed actions in equilibrium that will induce agent $n$ to choose action $1$ ($0$) when her private signal is $s_n$. We know that
\begin{align*}
&\mathcal{P}_{\sigma^*}(a_n=\theta|s_n)\\
=&\frac{f_0(s_n)\mathcal{P}_{\sigma^*}(h_n(\sigma^{*1}_n(s_n))\in H_n^0(s_n)|\theta=0)+f_1(s_n)\mathcal{P}_{\sigma^*}(h_n(\sigma^{*1}_n(s_n))\in H_n^1(s_n)|\theta=1)}{f_0(s_n)+f_1(s_n)}\\
=&\frac{f_0(s_n)}{f_0(s_n)+f_1(s_n)}\mathcal{P}_{\sigma^*}(h_n(\sigma^{*1}_n(s_n))\in H_n^0(s_n)|\theta=0)\\
&+(1-\frac{f_0(s_n)}{f_0(s_n)+f_1(s_n)})\mathcal{P}_{\sigma^*}(h_n(\sigma^{*1}_n(s_n))\in H_n^1(s_n)|\theta=1).
\end{align*}
Hence, the marginal benefit of observation is
\begin{align*}
&\mathcal{P}_{\sigma^*}(a_n=\theta|s_n)-\frac{f_1(s_n)}{f_0(s_n)+f_1(s_n)}\\
=&\frac{f_0(s_n)}{f_0(s_n)+f_1(s_n)}\mathcal{P}_{\sigma^*}(h_n(\sigma^{*1}_n(s_n))\in H_n^0(s_n)|\theta=0)\\
&-\frac{f_1(s_n)}{f_0(s_n)+f_1(s_n)}\mathcal{P}_{\sigma^*}(h_n(\sigma^{*1}_n(s_n))\in H_n^0(s_n)|\theta=1).
\end{align*}

Now, consider any $s_n^1>s_n^2\geq 0$, and the following sub-optimal strategy $\sigma'_n(s_n^2)$ for agent $n$ when her private signal is $s_n^2$: observe the same neighborhood and given any observation, choose the same action as if her signal were $s_n^1$. The marginal benefit of observation under this strategy is
\begin{align*}
&\frac{f_0(s_n^2)}{f_0(s_n^2)+f_1(s_n^2)}\mathcal{P}_{\sigma^*}(h_n(\sigma^{*1}_n(s_n^1))\in H_n^0(s_n^1)|\theta=0)\\
&-\frac{f_1(s_n^2)}{f_0(s_n^2)+f_1(s_n^2)}\mathcal{P}_{\sigma^*}(h_n(\sigma^{*1}_n(s_n^1))\in H_n^0(s_n^1)|\theta=1).
\end{align*}

Because $\sigma_n^{*1}(s_n^1)\neq\varnothing$ by assumption, we know that

\begin{align*}
&\frac{f_0(s_n^1)}{f_0(s_n^1)+f_1(s_n^1)}\mathcal{P}_{\sigma^*}(h_n(\sigma^{*1}_n(s_n^1))\in H_n^0(s_n^1)|\theta=0)\\
&-\frac{f_1(s_n^1)}{f_0(s_n^1)+f_1(s_n^1)}\mathcal{P}_{\sigma^*}(h_n(\sigma^{*1}_n(s_n^1))\in H_n^0(s_n^1)|\theta=1)\geq c.
\end{align*}

By the MLRP, $\frac{f_1(s_n^2)}{f_0(s_n^2)+f_1(s_n^2)}<\frac{f_1(s_n^1)}{f_0(s_n^1)+f_1(s_n^1)}$ and $\frac{f_0(s_n^2)}{f_0(s_n^2)+f_1(s_n^2)}>\frac{f_0(s_n^1)}{f_0(s_n^1)+f_1(s_n^1)}$. Therefore, we have
\begin{align*}
&\frac{f_0(s_n^2)}{f_0(s_n^2)+f_1(s_n^2)}\mathcal{P}_{\sigma^*}(h_n(\sigma^{*1}_n(s_n^1))\in H_n^0(s_n^1)|\theta=0)\\
&-\frac{f_1(s_n^2)}{f_0(s_n^2)+f_1(s_n^2)}\mathcal{P}_{\sigma^*}(h_n(\sigma^{*1}_n(s_n^1))\in H_n^0(s_n^1)|\theta=1)>c,
\end{align*}
which implies that $\sigma_n^{*1}(s_n^2)\neq\varnothing$.

2: Consider any $s_n^1>s_n^2\geq 0$, and the following sub-optimal strategy $\sigma'_n(s_n^1)$ for agent $n$ when her private signal is $s_n^1$: observe the same neighborhood and, given any observation, choose the same action as if her signal were $s_n^2$. We have
\begin{align*}
&\mathcal{P}_{\sigma^*}(a_n=\theta|s_n^1)\geq\mathcal{P}_{\sigma^*_{-n},\sigma'_n(s_n^1)}(a_n=\theta|s_n^1)\\
=&\frac{f_0(s_n^1)\mathcal{P}_{\sigma^*}(h_n(\sigma^{*1}_n(s_n^2))\in H_n^0(s_n^2)|\theta=0)+f_1(s_n^1)\mathcal{P}_{\sigma^*}(h_n(\sigma^{*1}_n(s_n^2))\in H_n^1(s_n^2)|\theta=1)}{f_0(s_n^1)+f_1(s_n^1)}\\
=&\frac{f_0(s_n^1)}{f_0(s_n^1)+f_1(s_n^1)}\mathcal{P}_{\sigma^*}(h_n(\sigma^{*1}_n(s_n^2))\in H_n^0(s_n^2)|\theta=0)\\
&+(1-\frac{f_0(s_n^1)}{f_0(s_n^1)+f_1(s_n^1)})\mathcal{P}_{\sigma^*}(h_n(\sigma^{*1}_n(s_n^2))\in H_n^1(s_n^2)|\theta=1).
\end{align*}

Therefore, we know that
\begin{align*}
&\mathcal{P}_{\sigma^*}(a_n=\theta|s_n^1)-\mathcal{P}_{\sigma^*}(a_n=\theta|s_n^2)\\
\geq & \mathcal{P}_{\sigma^*_{-n},\sigma'^{1}_n(s_n^1)}(a_n=\theta|s_n^1)-\mathcal{P}_{\sigma^*}(a_n=\theta|s_n^2)\\
=& (\frac{f_0(s_n^2)}{f_0(s_n^2)+f_1(s_n^2)}-\frac{f_0(s_n^1)}{f_0(s_n^1)+f_1(s_n^1)})\\
&(\mathcal{P}_{\sigma^*}(h_n(\sigma^{*1}_n(s_n^2))\in H_n^1(s_n^2)|\theta=1)-\mathcal{P}_{\sigma^*}(h_n(\sigma^{*1}_n(s_n^2))\in H_n^0(s_n^2)|\theta=0)).
\end{align*}

Consider any $h\in H_n^0(s_n^2)$, and consider $h'$ from the same neighborhood such that every action $0$ ($1$) in $h$ is replaced by $1$ ($0$) in $h'$. We know from the definition of $H_n^0(s_n^2)$ that $f_0(s_n^2)\mathcal{P}_{\sigma^*}(h|\theta=0)>f_1(s_n^2)\mathcal{P}_{\sigma^*}(h|\theta=1)$; by the assumption that $s_n^2\geq 0$, we have $\mathcal{P}_{\sigma^*}(h|\theta=0)>\mathcal{P}_{\sigma^*}(h|\theta=1)$. By symmetry, it follows that $\mathcal{P}_{\sigma^*}(h'|\theta=1)=\mathcal{P}_{\sigma^*}(h|\theta=0)>\mathcal{P}_{\sigma^*}(h|\theta=1)=\mathcal{P}_{\sigma^*}(h'|\theta=0)$. Hence, we have $f_1(s_n^2)\mathcal{P}_{\sigma^*}(h'|\theta=1)>f_0(s_n^2)\mathcal{P}_{\sigma^*}(h'|\theta=0)$, i.e., $h'\in H_n^1(s_n^2)$. It then follows that $\mathcal{P}_{\sigma^*}(h_n(\sigma^{*1}_n(s_n^2))\in H_n^1(s_n^2)|\theta=1)\geq \mathcal{P}_{\sigma^*}(h_n(\sigma^{*1}_n(s_n^2))\in H_n^0(s_n^2)|\theta=0)$, which immediately implies that $\mathcal{P}_{\sigma^*}(a_n=\theta|s_n^1)\geq\mathcal{P}_{\sigma^*}(a_n=\theta|s_n^2)$.

3: This result follows directly from 1.
\end{proof}

\begin{proof}[Proof of Theorem 1]

First, note that when $c=0$, it is always feasible for an agent to observe and imitate her immediate predecessor, which guarantees her the same expected payoff as this immediate predecessor. Hence, we know that $\mathcal{P}_{\sigma^*}(a_n=\theta)$ is weakly increasing in $n$. Since this probability is upper bounded by $1$, the sequence $\{\mathcal{P}_{\sigma^*}(a_n=\theta)\}$ converges. Let $r$ denote the limit.

Suppose that $r<1$. Thus, for any $\epsilon>0$, $N$ exists such that for any $n>N$, $\mathcal{P}_{\sigma^*}(a_n=\theta)\in(r-\epsilon,r]$. Take one such $n$, and consider agent $n+1$ and her sub-optimal strategy of observing agent $n$. By the assumption of unbounded private belief, $\bar{s}_{n+1}$ and $\underline{s}_{n+1}$ exists such that agent $n+1$ is indifferent between following her own signal and following agent $n$'s action, i.e., the following two conditions are satisfied:
\begin{align*}
\frac{f_1(\bar{s}_{n+1})}{f_0(\bar{s}_{n+1})+f_1(\bar{s}_{n+1})}&=\frac{f_1(\bar{s}_{n+1})}{f_0(\bar{s}_{n+1})+f_1(\bar{s}_{n+1})}\mathcal{P}_{\sigma^*}(a_n=1|\theta=1)\\
&+\frac{f_0(\bar{s}_{n+1})}{f_0(\bar{s}_{n+1})+f_1(\bar{s}_{n+1})}\mathcal{P}_{\sigma^*}(a_n=0|\theta=0)\\
\frac{f_0(\underline{s}_{n+1})}{f_0(\underline{s}_{n+1})+f_1(\underline{s}_{n+1})}&=\frac{f_1(\underline{s}_{n+1})}{f_0(\underline{s}_{n+1})+f_1(\underline{s}_{n+1})}\mathcal{P}_{\sigma^*}(a_n=1|\theta=1)\\
&+\frac{f_0(\underline{s}_{n+1})}{f_0(\underline{s}_{n+1})+f_1(\underline{s}_{n+1})}\mathcal{P}_{\sigma^*}(a_n=0|\theta=0).
\end{align*}
The above equation can be further simplified as
\begin{align*}
\frac{f_1(\bar{s}_{n+1})}{f_0(\bar{s}_{n+1})}&=\frac{\mathcal{P}_{\sigma^*}(a_n=0|\theta=0)}{\mathcal{P}_{\sigma^*}(a_n=0|\theta=1)}\\
\frac{f_0(\underline{s}_{n+1})}{f_1(\underline{s}_{n+1})}&=\frac{\mathcal{P}_{\sigma^*}(a_n=1|\theta=1)}{\mathcal{P}_{\sigma^*}(a_n=1|\theta=0)}.
\end{align*}
By the previous argument, $\mathcal{P}_{\sigma^*}(a_n=\theta)=\frac{1}{2}\mathcal{P}_{\sigma^*}(a_n=1|\theta=1)+\frac{1}{2}\mathcal{P}_{\sigma^*}(a_n=0|\theta=0)\leq r$, and hence $\min\{\mathcal{P}_{\sigma^*}(a_n=1|\theta=1),\mathcal{P}_{\sigma^*}(a_n=0|\theta=0)\}\leq r$. Without loss of generality, assume that $\mathcal{P}_{\sigma^*}(a_n=1|\theta=1)\leq r$. Then, $\mathcal{P}_{\sigma^*}(a_n=0|\theta=1)\geq 1-r$, and hence $\frac{f_1(\bar{s}_{n+1})}{f_0(\bar{s}_{n+1})}\leq\frac{1}{1-r}$. Let $\hat{s}$ be the value of the private signal such that $\frac{f_1(\hat{s})}{f_0(\hat{s})}=\frac{2}{1-r}$. We know that
\begin{align*}
\mathcal{P}_{\sigma^*}(a_{n+1}=\theta)-\mathcal{P}_{\sigma^*}(a_{n}=\theta)&\geq\int_{\hat{s}}^1 \frac{1}{2}(f_1(s)\mathcal{P}_{\sigma^*}(a_n=0|\theta=1)-f_0(s)\mathcal{P}_{\sigma^*}(a_n=0|\theta=0)) ds\\
&\geq \int_{\hat{s}}^1 \frac{1}{2}((1-r)f_1(s)-f_0(s))ds\\
&\geq \int_{\hat{s}}^1 \frac{1}{2}f_0(s)ds.
\end{align*}
Therefore, we have
\begin{align*}
\int_{\hat{s}}^1 \frac{1}{2}f_0(s)ds\leq \mathcal{P}_{\sigma^*}(a_{n+1}=\theta)-\mathcal{P}_{\sigma^*}(a_{n}=\theta)<\epsilon.
\end{align*}
For sufficiently small $\epsilon$, this inequality is violated, and thus we have a contradiction.
\end{proof}

\begin{proof}[Proof of Corollary 1]
By Theorem 1 in Acemoglu et al. (2011), any network topology that does not have expanding observations cannot support asymptotic learning. Because asymptotic learning occurs in every equilibrium when $c=0$ and agents have unbounded private beliefs, it must follow that $\mathcal{Q}_{\sigma^*}$ has expanding observation for any $\sigma^*$.
\end{proof}

\begin{proof}[Proof of Proposition 3]
Part ($a$) is already proved by Smith and Sorensen\cite{SS} in their Theorem 1. Acemoglu et al.\cite{ADLO} offer an alternative proof in their Theorem 3. First, note that when the cost of observation is zero and $K(n)=n-1$ for all $n$, for every agent with any private signal in any equilibrium, observing her optimal choice of neighborhood generates the same equilibrium behavior as in the equilibrium in which each agent observes all her predecessors' actions regardless of her private signal because any action that the agent chooses not to observe must have no influence on her own action. Second, in the equilibrium where each agent observes all her predecessors' actions regardless of her private signal, the agents' behavior coincides with that in a model where this observation structure (each agent observing the entire action history before her) is exogenously given. Hence, the proofs above apply directly.

For Part ($b$), assume that $c=0$ and private beliefs are bounded, and suppose that there is an equilibrium in which asymptotic learning occurs. I first show that for all $M\in\mathbb{N}$, $N\in\mathbb{N}$ exists such that $\max_{b\in\sigma_n^{*1}(s_n)}b>M$ for every $s_n$ and every $n>N$.

To prove the above claim, first note that by assumption, $N'\in\mathbb{N}$ exists such that $K(n)>0$ for all $n>N'$. It immediately follows from $c=0$ that $\mathcal{P}_{\sigma^*}(a_n=\theta)$ is weakly increasing in $n$ for all $n>N'$. Next, for any $M\in\mathbb{N}$, let $\hat{a}(s)$ denote the action that maximizes the expected payoff for an agent whose private signal is $s$ and who observes the neighborhood $\{1,\cdots,M\}$. It is clear that $\sup_{s\in(-1,1)}\mathcal{P}_{\sigma^*_1,\cdots,\sigma^*_M}(\hat{a}(s)=\theta)<1$ because private beliefs are bounded and $M$ is finite. Because asymptotic learning occurs by assumption, $N''\in\mathbb{N}$ exists such that $\mathcal{P}_{\sigma^*}(a_n=\theta)>\sup_{s\in(-1,1)}\mathcal{P}_{\sigma^*_1,\cdots,\sigma^*_M}(\hat{a}(s)=\theta)$ for all $n>N''$. Let $N=\max\{N',N''\}+1$. For any agent $n>N$, by observing agent $N$ and copying agent $N$'s action, she can achieve a strictly higher probability of taking the right action than by observing $\{1,\cdots,M\}$. Hence, it must be the case that $\max_{b\in\sigma_n^{*1}(s_n)}b>M$ for every $s_n$ and every $n>N$. Then, the argument in the proof of Theorem 3 of Acemoglu et al.\cite{ADLO}, whose validity requires only the existence of capacity upper bound $\bar{K}$, can be applied to show that asymptotic learning does not occur in any equilibrium.
\end{proof}

\begin{proof}[Proof of Lemma 1]
The definition of $s^*$ implies that when agent $n$ has a private signal of $s^*$, he is indifferent between paying $c$ to know the true state and choosing accordingly, and paying nothing and choosing $1$. Note that the largest possible benefit from observing is always strictly less than knowing the true state with certainty. Hence, the (positive) private signal that makes agent $n$ indifferent between observing and not observing must be less than $s^*$.
\end{proof}

\begin{proof}[Proof of Lemma 2]
Suppose that when the capacity structure has finite observations, maximal learning occurs in some equilibrium $\sigma^*$. It follows that $K\in\mathbb{N}$ exists such that, for any $\epsilon>0$ and $N\in\mathbb{N}$, $n>N$ exists such that $\mathcal{P}_{\sigma^*}(a_n=\theta|s_n\in(-s_*^n,s_*^n))>1-\epsilon$ and $0<K(n)\leq K$.

Consider one such agent $n$. For at least one state $\theta\in(0,1)$, there must be some $s_n\in(-s_*^n,s_*^n)$ such that $\mathcal{P}_{\sigma^*}(a_n=\theta|\theta,s_n)>1-\epsilon$. Without loss of generality, assume that (one) such $\theta$ is $1$. Since $s_n\in(-s_*^n,s_*^n)$, there must be some realized action sequence in $n$'s observed neighborhood $\sigma^{*1}_n(s_n)$, given which $n$ will take action $0$. Because, by Lemma 1, $K(n)\leq K$, we know that when the true state is $1$, the probability of this action sequence occurring is bounded below by $\min\{F_1(s_*),1-F_1(s^*)\}^K$. Hence, we have
\begin{align*}
1-\epsilon<\mathcal{P}_{\sigma^*}(a_n=1|1,s_n)\leq 1-\min\{F_1(s_*),1-F_1(s^*)\}^K
\end{align*}
However, for sufficiently small $\epsilon$ we have $1-\epsilon>1-\min\{F_1(s_*),1-F_1(s^*)\}^K$, which is a contradiction.
\end{proof}

\begin{proof}[Proof of Lemma 3]
I prove here that $\lim_{\epsilon\rightarrow 0^+}(\lim\sup_{k\rightarrow\infty}\mathcal{P}_{\sigma^*}(R_{\sigma^*}^{B_k}>1-\epsilon|0,s))=0$, and the second equality would follow from an analogous argument. Suppose the equality does not hold, then $s\in(s_*,s^*)$ and $\rho>0$ exist such that for any $\epsilon>0$ and any $N\in\mathbb{N}$, $k>N$ exists such that $\mathcal{P}_{\sigma^*}(R_{\sigma^*}^{B_k}>1-\epsilon|0,s)>\rho$. Consider any realized decision sequence $h_{\epsilon}$ from $B_k$ that, together with $s$, induces some $r>1-\epsilon$, and let $H_{\epsilon}$ denote the set of all such decision sequences; thus, we know that
\begin{align*}
\frac{\mathcal{P}_{\sigma^*}(h_{\epsilon}|\theta')f_{\theta'}(s)}{\mathcal{P}_{\sigma^*}(h_{\epsilon}|\theta)f_{\theta}(s)+\mathcal{P}_{\sigma^*}(h_{\epsilon}|\theta')f_{\theta'}(s)}&=r\\
\sum_{h_{\epsilon}\in H_{\epsilon}}\mathcal{P}_{\sigma^*}(h_{\epsilon}|\theta)&>\rho.
\end{align*}
The above two conditions imply that
\begin{align*}
1\geq \sum_{h_{\epsilon}\in H_{\epsilon}}\mathcal{P}_{\sigma^*}(h_{\epsilon}|\theta')>\frac{(1-\epsilon)\rho f_{\theta}(s)}{\epsilon f_{\theta'}(s)}.
\end{align*}
For sufficiently small $\epsilon$, we have $\frac{(1-\epsilon)\rho f_{\theta}(s)}{\epsilon f_{\theta'}(s)}>1$, which is a contradiction.
\end{proof}

\begin{proof}[Proof of Lemma 4]
Without loss of generality, assume that $\hat{r}\in(0,r)$. We know that there is a private signal $s$ and an action sequence $h$ from $B_k$ such that
\begin{align*}
r=\frac{\mathcal{P}_{\sigma^*}(h|1)f_1(s)}{\mathcal{P}_{\sigma^*}(h|1)f_1(s)+\mathcal{P}_{\sigma^*}(h|0)f_0(s)}.
\end{align*}
Consider $h\cup\{a_{k+1}\}$ where $a_{k+1}=0$. The new belief would then be
\begin{align*}
r_1=\frac{\mathcal{P}_{\sigma^*}(h|1)f_1(s)\times \mathcal{P}_{\sigma^*}(a_{k+1}=0|h,1)}{\mathcal{P}_{\sigma^*}(h|1)f_1(s)\times \mathcal{P}_{\sigma^*}(a_{k+1}=0|h,1)+\mathcal{P}_{\sigma^*}(h|0)f_0(s)\times \mathcal{P}_{\sigma^*}(a_{k+1}=0|h,0)}.
\end{align*}
Note that
\begin{align*}
\mathcal{P}_{\sigma^*}(a_{k+1}=0|h,1)&=F_1(-s_*^{k+1})+\mathcal{P}_{\sigma^*}(a_{k+1}=0,\text{ observe}|h,1)\\
\mathcal{P}_{\sigma^*}(a_{k+1}=0|h,0)&=F_0(-s_*^{k+1})+\mathcal{P}_{\sigma^*}(a_{k+1}=0,\text{ observe}|h,0),
\end{align*}
and that
\begin{align*}
\mathcal{P}_{\sigma^*}(a_{k+1}=0,\text{ observe}|h,1)&=\int_{-s_*^{k+1}}^{s_*^{k+1}}\mathcal{P}_{\sigma^*}(a_{k+1}=0,|h,1,s_{k+1})f_1(s_{k+1})ds_{k+1}\\
\mathcal{P}_{\sigma^*}(a_{k+1}=0,\text{ observe}|h,0)&=\int_{-s_*^{k+1}}^{s_*^{k+1}}\mathcal{P}_{\sigma^*}(a_{k+1}=0,|h,0,s_{k+1})f_0(s_{k+1})ds_{k+1}.
\end{align*}
In any equilibrium, note that $\mathcal{P}_{\sigma^*}(a_{k+1}=0,|h,1,s_{k+1})=\mathcal{P}_{\sigma^*}(a_{k+1}=0,|h,0,s_{k+1})$ for any given $h$ and $s_{k+1}\in[-s_*^{k+1},s_*^{k+1}]$. Moreover, given any $s_{k+1}\in[0,s_*^{k+1}]$, $\mathcal{P}_{\sigma^*}(a_{k+1}=0,|h,1,s_{k+1})$ and $\mathcal{P}_{\sigma^*}(a_{k+1}=0,|h,0,s_{k+1})$ are either $0$ or $1$.

For any $s_{k+1}\in[0,s_*^{k+1}]$, note that in a symmetric equilibrium, agent $k+1$ observes the same neighborhood, given private signal $s_{k+1}$ and $-s_{k+1}$. Hence, if $k+1$ chooses $1$ with private signal $-s_{k+1}$, then he will also choose $1$ with private signal $s_{k+1}$; if $k+1$ chooses $0$ with private signal $s_{k+1}$, then he will also choose $0$ with private signal $-s_{k+1}$. Together with the assumptions of symmetric signal structure and the MLRP, which imply that $f_1(-s_{k+1})=f_0(s_{k+1})\leq f_1(s_{k+1})=f_0(-s_{k+1})$, it then follows that
\begin{align*}
&\mathcal{P}_{\sigma^*}(a_{k+1}=0,|h,1,s_{k+1})f_1(s_{k+1})+\mathcal{P}_{\sigma^*}(a_{k+1}=0,|h,1,-s_{k+1})f_1(-s_{k+1})\\
\leq&\mathcal{P}_{\sigma^*}(a_{k+1}=0,|h,0,s_{k+1})f_0(s_{k+1})+\mathcal{P}_{\sigma^*}(a_{k+1}=0,|h,0,-s_{k+1})f_0(-s_{k+1}).
\end{align*}
Therefore, we have $\mathcal{P}_{\sigma^*}(a_{k+1}=0,\text{ observe}|h,0)\geq \mathcal{P}_{\sigma^*}(a_{k+1}=0,\text{ observe}|h,1)$. Together with Lemma 1, we have
\begin{align*}
\frac{\mathcal{P}_{\sigma^*}(a_{k+1}=0|h,1)}{\mathcal{P}_{\sigma^*}(a_{k+1}=0|h,0)}\leq\frac{F_1(s_*^{k+1})}{F_0(s_*^{k+1})}<\frac{F_1(s^*)}{F_0(s^*)}<1.
\end{align*}
The second inequality is based on the fact that $F_1(s^*)-F_1(-s^*)=F_0(s^*)-F_0(-s^*)$ by the symmetry of the signal structure. Therefore, we have
\begin{align*}
\frac{r}{r_1}=\frac{1+\frac{\mathcal{P}_{\sigma^*}(h|0)f_0(s)}{\mathcal{P}_{\sigma^*}(h|1)f_1(s)}\frac{\mathcal{P}_{\sigma^*}(a_{k+1}=0|h,0)}{\mathcal{P}_{\sigma^*}(a_{k+1}=0|h,1)}}{1+\frac{\mathcal{P}_{\sigma^*}(h|0)f_0(s)}{\mathcal{P}_{\sigma^*}(h|1)f_1(s)}}=r+(1-r)\frac{\mathcal{P}_{\sigma^*}(a_{k+1}=0|h,0)}{\mathcal{P}_{\sigma^*}(a_{k+1}=0|h,1)}>r+(1-r)\frac{F_0(s^*)}{F_1(s^*)}.
\end{align*}

Note that the expression on the right-hand side above is decreasing in $r$. Let $r_m$ denote the belief induced by $h\cup\{a_{k+1},\cdots, a_{k+m}\}$ where $a_{k+1}=\cdots=a_{k+m}=0$. We have
\begin{align*}
r_m=r\times\frac{r_1}{r}\times\cdots\times\frac{r_m}{r_{m-1}}<r\times(\frac{r_1}{r})^m.
\end{align*}

Because $\frac{r_1}{r}=\frac{1}{r+(1-r)\frac{F_0(s^*)}{F_1(s^*)}}<1$, we can find the desired $N(r,\hat{r})$ for any $\hat{r}\in(0,r)$, such that a realized belief that is less than $\hat{r}$ can be induced by $s$ and $h\cup\{a_{k+1},\cdots,a_{k+{N(r,\hat{r})}}\}$, where $a_{k+1}=\cdots=a_{k+{N(r,\hat{r})}}=0$.
\end{proof}

\begin{proof}[Proof of Lemma 5]
Suppose not, then noting that $\mathcal{P}_{\sigma^*}^{B_k}(\hat{a}\neq \theta|s)$ must be weakly decreasing in $k$, it follows that $\lim_{k\rightarrow\infty}\mathcal{P}_{\sigma^*}^{B_k}(\hat{a}\neq \theta|s)>0$. Let $\rho>0$ denote this limit. From Lemma 3, we know that for any $\alpha>0$ and for either true state $\theta=0,1$, $z\in[\frac{1}{2},1)$ exists such that $M\in\mathbb{N}$ exists such that $\max\{\mathcal{P}_{\sigma^*}(R_{\sigma^*}^{B_{k}}>z|0,s),\mathcal{P}_{\sigma^*}(1-R_{\sigma^*}^{B_{k}}>z|1,s)\}<\alpha$ for any $k>M$. Let $\alpha=\frac{1}{2}\rho$, then we have $\max\{\mathcal{P}_{\sigma^*}(R_{\sigma^*}^{B_k}>z|0,s),\mathcal{P}_{\sigma^*}(1-R_{\sigma^*}^{B_k}>z|1,s)\}<\frac{1}{2}\rho$ for any $k>M$. Then, for any $\delta>0$, we can find a sufficiently large $k$ such that for any $k'\geq k$, (1) $\mathcal{P}_{\sigma^*}^{B_{k'}}(\hat{a}\neq \theta|s)\in(\rho,\rho+\delta)$ and (2) $\max\{\mathcal{P}_{\sigma^*}(R_{\sigma^*}^{B_{k'}}>z|0,s),\mathcal{P}_{\sigma^*}(1-R_{\sigma^*}^{B_{k'}}>z|1,s)\}<\frac{1}{2}\rho$. Hence, we have
\begin{align*}
&\frac{f_0(s)}{f_0(s)+f_1(s)}\mathcal{P}_{\sigma^*}(R_{\sigma^*}^{B_{k'}}\in[\frac{1}{2},z]|0,s)+\frac{f_1(s)}{f_0(s)+f_1(s)}\mathcal{P}_{\sigma^*}(1-R_{\sigma^*}^{B_{k'}}\in[\frac{1}{2},z]|1,s)\\
=&\mathcal{P}_{\sigma^*}^{B_{k'}}(\hat{a}\neq \theta|s)-\frac{f_0(s)}{f_0(s)+f_1(s)}\mathcal{P}_{\sigma^*}(R_{\sigma^*}^{B_{k'}}>z|0,s)-\frac{f_1(s)}{f_0(s)+f_1(s)}\mathcal{P}_{\sigma^*}(1-R_{\sigma^*}^{B_{k'}}>z|1,s)>\frac{1}{2}\rho
\end{align*}

By Lemma 3, for any $\pi>0$, $N(\pi)=\max\{N(z,\frac{1}{2+\pi}),N(1-z,1-\frac{1}{2+\pi})\}\in\mathbb{N}$ exists such that whenever $\theta=0$ and $R_{\sigma^*}^{B_k}\in[\frac{1}{2},z]$ or $\theta=1$ and $1-R_{\sigma^*}^{B_k}\in[\frac{1}{2},z]$, additional $N(\pi)$ observations can reverse an incorrect decision. Consider the following (sub-optimal) updating method for a rational agent who observes $B_{k'}=B_{k+N(\pi)}$: switch her action from $1$ to $0$ if and only if $R_{\sigma^*}^{B_k}\in[\frac{1}{2},z]$, and $a_{k+1}=\cdots=a_{k+N(\pi)}=0$; switch her action from $0$ to $1$ if and only if $1-R_{\sigma^*}^{B_k}\in[\frac{1}{2},z]$, and $a_{k+1}=\cdots=a_{k+N(\pi)}=1$. Let $h$ denote a decision sequence from $B_k$ that, together with $s$, induces such a posterior belief in the former case, and let $h'$ denote a decision sequence from $B_k$ that, together with $s$, induces such a posterior belief in the latter case. Let $H$ and $H'$ denote the sets of such decision sequences correspondingly. We have
\begin{align*}
&\mathcal{P}_{\sigma^*}^{B_k}(\hat{a}\neq \theta|s)-\mathcal{P}_{\sigma^*}^{B_{k'}}(\hat{a}\neq \theta|s)\\
\geq&\sum_{h\in H}(\frac{f_0(s)}{f_0(s)+f_1(s)}\mathcal{P}_{\sigma^*}(h,a_{k+1}=\cdots=a_{k+N(\pi)}=0|0)\\
&-\frac{f_1(s)}{f_0(s)+f_1(s)}\mathcal{P}_{\sigma^*}(h,a_{k+1}=\cdots=a_{k+N(\pi)}=0|1))\\
&+\sum_{h'\in H'}(\frac{f_1(s)}{f_0(s)+f_1(s)}\mathcal{P}_{\sigma^*}(h',a_{k+1}=\cdots=a_{k+N(\pi)}=1|1)\\
&-\frac{f_0(s)}{f_0(s)+f_1(s)}\mathcal{P}_{\sigma^*}(h',a_{k+1}=\cdots=a_{k+N(\pi)}=1|0)).
\end{align*}
From the proof of Lemma 4, we know that for every $h$,
\begin{align*}
\frac{\mathcal{P}_{\sigma^*}(h,a_{k+1}=\cdots=a_{k+N(\pi)}=0|0)f_0(s)}{\mathcal{P}_{\sigma^*}(h,a_{k+1}=\cdots=a_{k+N(\pi)}=0|0)f_0(s)+\mathcal{P}_{\sigma^*}(h,a_{k+1}=\cdots=a_{k+N(\pi)}=0|1)f_1(s)}\geq\frac{1+\pi}{2+\pi},
\end{align*}
which implies that
\begin{align*}
&\mathcal{P}_{\sigma^*}(h,a_{k+1}=\cdots=a_{k+N(\pi)}=0|0)f_0(s)-\mathcal{P}_{\sigma^*}(h,a_{k+1}=\cdots=a_{k+N(\pi)}=0|1)f_1(s)\\
\geq&\pi f_1(s)\mathcal{P}_{\sigma^*}(h,a_{k+1}=\cdots=a_{k+N(\pi)}=0|1)\\
\geq&\pi f_1(s)F_1(-s^*)^{N(\pi)}\mathcal{P}_{\sigma^*}(h|1).
\end{align*}
By the definition of $h$, we have
\begin{align*}
\frac{1}{2}\leq\frac{\mathcal{P}_{\sigma^*}(h|1)f_1(s)}{\mathcal{P}_{\sigma^*}(h|1)f_1(s)+\mathcal{P}_{\sigma^*}(h|0)f_0(s)}\leq z,
\end{align*}
which implies that
\begin{align*}
\mathcal{P}_{\sigma^*}(h|1)f_1(s)\geq \mathcal{P}_{\sigma^*}(h|0)f_0(s).
\end{align*}
Similarly, we have
\begin{align*}
&\mathcal{P}_{\sigma^*}(h',a_{k+1}=\cdots=a_{k+N(\pi)}=1|1)f_1(s)-\mathcal{P}_{\sigma^*}(h',a_{k+1}=\cdots=a_{k+N(\pi)}=1|0)f_0(s)\\
\geq&\pi f_0(s)(1-F_0(s^*))^{N(\pi)}\mathcal{P}_{\sigma^*}(h'|0)=\pi f_0(s)F_1(-s^*)^{N(\pi)}\mathcal{P}_{\sigma^*}(h'|0),
\end{align*}
and
\begin{align*}
\mathcal{P}_{\sigma^*}(h'|0)f_0(s)\geq \mathcal{P}_{\sigma^*}(h'|1)f_1(s).
\end{align*}
From the previous construction, we know that
\begin{align*}
&\frac{f_0(s)}{f_0(s)+f_1(s)}\sum_{h\in H} \mathcal{P}_{\sigma^*}(h|0)+\frac{f_1(s)}{f_0(s)+f_1(s)}\sum_{h'\in H'} \mathcal{P}_{\sigma^*}(h'|1)\\
=&\frac{f_0(s)}{f_0(s)+f_1(s)}\mathcal{P}_{\sigma^*}(R_{\sigma^*}^{B_{k'}}\in[\frac{1}{2},z]|0,s)+\frac{f_1(s)}{f_0(s)+f_1(s)}\mathcal{P}_{\sigma^*}(1-R_{\sigma^*}^{B_{k'}}\in[\frac{1}{2},z]|1,s)>\frac{1}{2}\rho.
\end{align*}
Combining the previous inequalities, we have
\begin{align*}
&\mathcal{P}_{\sigma^*}^{B_k}(\hat{a}\neq \theta|s)-\mathcal{P}_{\sigma^*}^{B_{k'}}(\hat{a}\neq \theta|s)\\
>&\pi F_1(-s^*)^{N(\pi)}\frac{1}{2}\rho.
\end{align*}
From the previous construction, we also know that
\begin{align*}
\mathcal{P}_{\sigma^*}^{B_k}(\hat{a}\neq \theta|s)-\mathcal{P}_{\sigma^*}^{B_{k'}}(\hat{a}\neq \theta|s)<\delta.
\end{align*}
Clearly, for some given $\pi>0$, a sufficiently small $\delta$ exists such that $\pi F_1(-s^*)^{N(\pi)}\frac{1}{2}\rho>\delta$, which implies a contradiction.
\end{proof}

\begin{proof}[Proof of Theorem 2]
Lemma 2 has already shown that maximal learning occurs in every equilibrium only if the capacity structure has infinite observations. Now I prove the sufficiency of infinite observations for maximal learning in every equilibrium. Note that in any equilibrium $\sigma^*$, for any $n$ and any $s_n\in(-s_*^n,s_*^n)$,
\begin{align*}
\mathcal{P}_{\sigma^*}(a_n=\theta|s_n)\geq \mathcal{P}_{\sigma^*}^{B_{K(n)}}(a_n=\theta|s_n).
\end{align*}
When the capacity structure has infinite observations, i.e., $\lim_{n\rightarrow\infty}K(n)=\infty$, by Lemma 5 we know that
\begin{align*}
\lim_{n\rightarrow\infty}\mathcal{P}_{\sigma^*}^{B_{K(n)}}(a_n=\theta|s_n)=1,
\end{align*}
and thus
\begin{align*}
\lim_{n\rightarrow\infty}\mathcal{P}_{\sigma^*}(a_n=\theta|s_n)=1,
\end{align*}
which further implies that
\begin{align*}
\lim_{n\rightarrow\infty}\mathcal{P}_{\sigma^*}(a_n=\theta)=\frac{1}{2}F_0(s_*)+\frac{1}{2}(1-F_1(s_*))=P^*(c).
\end{align*}
Therefore, maximal learning occurs in every equilibrium.
\end{proof}

\begin{proof}[Proof of Corollary 2]
To apply the argument for Theorem 2 without the symmetry assumption, it suffices to show that for any agent $n$ and any observed action sequence $h$ in neighborhood $\{1,\cdots,n-1\}$, $\mathcal{P}_{\sigma^*}(a_{n}=0|h,0)-\mathcal{P}_{\sigma^*}(a_{n}=0,|h,1)$ and $\mathcal{P}_{\sigma^*}(a_{n}=1|h,1)-\mathcal{P}_{\sigma^*}(a_{n}=1,|h,0)$ are both positive and bounded away from zero.

Note that
\begin{align*}
\mathcal{P}_{\sigma^*}(a_{n}=0|h,0)&>F_0(s_*)\\
\mathcal{P}_{\sigma^*}(a_{n}=0|h,1)&<F_1(s^*)\\
\mathcal{P}_{\sigma^*}(a_{n}=1|h,1)&>1-F_1(s^*)\\
\mathcal{P}_{\sigma^*}(a_{n}=1|h,0)&<1-F_0(s_*).
\end{align*}

Hence, we have
\begin{align*}
\mathcal{P}_{\sigma^*}(a_{n}=0|h,0)-\mathcal{P}_{\sigma^*}(a_{n}=0,|h,1)&>F_0(s_*)-F_1(s^*)\\
\mathcal{P}_{\sigma^*}(a_{n}=1|h,1)-\mathcal{P}_{\sigma^*}(a_{n}=1,|h,0)&>F_0(s_*)-F_1(s^*).
\end{align*}

Therefore, the assumption $F_0(s_*)>F_1(s^*)$ suffices for the above two differences to be positive and bounded away from zero.
\end{proof}

\begin{proof}[Proof of Corollary 3]
Again, it suffices to show that for any agent $n$ with $K(n)=1$ or $K(n)=n-1$ and any observed action sequence $h$ in neighborhood $\{1,\cdots,n-1\}$, $\mathcal{P}_{\sigma^*}(a_{n}=0|h,0)-\mathcal{P}_{\sigma^*}(a_{n}=0,|h,1)$ and $\mathcal{P}_{\sigma^*}(a_{n}=1|h,1)-\mathcal{P}_{\sigma^*}(a_{n}=1,|h,0)$ are both positive and bounded away from zero.

Note that when $K(n)=1$ or $K(n)=n-1$, $n$'s observed neighborhood (whenever she chooses to observe) is the same regardless of her signal. When $K(n)=1$, she observes the predecessor with the largest probability of taking the correct action; when $K(n)=n-1$, she observes the entire set of predecessors. Hence, given any observed action sequence $h$ in neighborhood $\{1,\cdots,n-1\}$, there must be a signal $s'(h)\in (s_*,s^*)$ such that $n$ will choose action 1 if $s_n>s'$ and 0 if $s_n<s'$. Hence, by MLRP we know that
\begin{align*}
\mathcal{P}_{\sigma^*}(a_{n}=0|h,0)-\mathcal{P}_{\sigma^*}(a_{n}=0,|h,1)&=\mathcal{P}_{\sigma^*}(a_{n}=1|h,1)-\mathcal{P}_{\sigma^*}(a_{n}=1,|h,0)\\
&=F_0(s'(h))-F_1(s'(h))>F_0(s^*)-F_1(s^*)>0.
\end{align*}
Therefore, the above two differences are positive and bounded away from zero.
\end{proof}

\begin{proof}[Proof of Proposition 4]
Let $\bar{\beta}=\lim_{s\rightarrow 1}\frac{f_1(s)}{f_0(s)+f_1(s)}$ and $\underline{\beta}=\lim_{s\rightarrow -1}\frac{f_1(s)}{f_0(s)+f_1(s)}$. Let $\Delta=\frac{\underline{\beta}(1-\bar{\beta})}{-\underline{\beta}+\bar{\beta}+2\underline{\beta}(1-\bar{\beta})}\in(0,1)$. Acemoglu et al. (2011) have shown that when agents have bounded private beliefs and when each agent observes all her predecessors, the learning probability of any agent is bounded above by $\max\{\Delta,1-\Delta\}<1$, i.e., $P^*(\sigma^*(0))\leq\max\{\Delta,1-\Delta\}<1$.

Note that for any $\epsilon\in(0,1)$, $c'>0$ exists such that agents have strong private beliefs under $c'$, and that $F_0(s^*)-F_0(-s^*)=F_1(s^*)-F_1(-s^*)=1-\epsilon$ (due to symmetry and continuity of signal distributions). The capacity structure $K(n)=n-1$ has infinite observations; thus, by Theorem 2, maximal learning occurs in any equilibrium $\sigma^*(c')$, which implies that
\begin{align*}
\lim_{n\rightarrow\infty}\mathcal{P}_{\sigma^*(c')}(a_n=\theta)=F_0(s^*)-F_0(-s^*)+F_0(-s^*)=F_0(s^*)>1-\epsilon.
\end{align*}

Let $\epsilon=1-\max\{\Delta,1-\Delta\}\in(0,1)$, then $c'$ exists such that $P^*(\sigma^*(0))<P^*(\sigma^*(c'))$ in any equilibrium $\sigma^*(c')$. Again, by continuity of the signal distributions, the desired $\bar{c},\underline{c}$ exist.
\end{proof}

\begin{proof}[Proof of Proposition 5]
Let $(F_0,F_1)$ and $(G_0,G_1)$ denote two (symmetric) signal structures that both generate strong private beliefs, and let $s_F^*$ and $s_G^*$ be the positive private signals such that an agent is indifferent between paying $c$ to know the true state and not paying $c$ and acting according to the signal. Assume that $(F_0,F_1)$ has higher signal strength: $F_0(-s^*_F)+1-F_0(s^*_F)>G_0(-s^*_G)+1-G_0(s^*_G)$. We already know that for any equilibrium $\sigma^*_F$ under the first signal structure and for any equilibrium $\sigma^*_G$ under the second signal structure,
\begin{align*}
\lim_{n\rightarrow\infty}\mathcal{P}_{\sigma^*_F}(a_n=\theta)&=F_0(s^*_F)\\
\lim_{n\rightarrow\infty}\mathcal{P}_{\sigma^*_G}(a_n=\theta)&=G_0(s^*_G).
\end{align*}

Hence, when $G_0(s^*_G)>F_0(s^*_F)$, the signal structure with higher strength will lead to lower limit learning probability.
\end{proof}

\begin{proof}[Proof of Proposition 6]
Part (a) follows from Theorem 1. When $c(1)>0$, it is clear that asymptotic learning does not occur in any equilibrium. When $c(1)=0$, we know that in any equilibrium $\sigma^*$, the probability of taking the right action $\mathcal{P}_{\sigma^*}(a_n=\theta)$ is weakly increasing in $n$. Suppose that asymptotic learning does not occur in $\sigma^*$, then $\mathcal{P}_{\sigma^*}(a_n=\theta)$ must converge to a limit in $(0,1)$. Then the argument in the proof for Theorem 1 can be applied to derive a contradiction.

Part (b) follows from Theorem 2. By Theorem 2, maximal learning requires infinite observations as $n$ grows large. When $c(m+1)-c(m)=0$ for all $m\geq 1$, it is clear that maximal learning occurs in every equilibrium. When $c(m+1)-c(m)>0$ for some $m$, in any equilibrium, consider the case when an agent receives a private signal that makes her indifferent between not observing and paying $c(1)$ to know the true state. We know that such a private signal exists because by assumption, private beliefs are unbounded, signal distributions are continuous and $c(1)>0$. Since $c(m+1)-c(m)>0$ for some $m$, the cost of observing infinitely many actions is strictly higher than $c(1)$, and hence the agent will choose not to observe, regardless of her capacity of observation. Because signal distributions are continuous, there exists a positive measure of signals given which the agent will choose not to observe. Therefore, maximal learning does not occur in any equilibrium.
\end{proof}

\begin{proof}[Proof of Proposition 7]
``Only if'': assume that $n$ does not exist. Then no agent will observe and clearly asymptotic learning does not occur in any equilibrium.

``If'': the condition implies that agent $n$ will observe $K(n)$ of her predecessors. Then, beginning with $n$, observing will be weakly better than not observing for each agent because an agent can at least observe agent $n$ and achieve exactly the same expected payoff as agent $n$. Then, the argument for Theorem 1 can be applied to show that asymptotic learning occurs in every equilibrium.
\end{proof}

\begin{proof}[Proof of Proposition 8]
``Only if'': see the proof of Theorem 2.

``If'': since the capacity structure has infinite observations, we can construct a sequence of agents $\{i_m\}_{m=1}^{\infty}$ such that $K(i_m)\leq K(i_{m+1})$ for any $m$, and $\lim_{m\rightarrow \infty}K(i_m)=\infty$. It thus suffices to show that maximal learning occurs in this sequence of agents.

Denote $\bar{s}_*^{i_m}$ and $\underline{s}_*^{i_m}$ as the positive and negative private signals such that agent $i_m$ is indifferent between observing and not observing (when either of such private signals does not exist, let $\bar{s}_*^{i_m}$ or $\underline{s}_*^{i_m}$ be the private signal such that $i_m$ is indifferent between choosing $1$ and $0$). Clearly, both $\bar{s}_*^{i_m}$ and $\underline{s}_*^{i_m}$ converge. Denote $\bar{s}$ and $\underline{s}$ as the corresponding limits. We know that for any $m$, $F_0(\underline{s}_*^{i_m})-F_1(\underline{s}_*^{i_m})\geq F_0(\underline{s})-F_1(\underline{s})>0$. Let $\epsilon=\frac{1}{2}(F_0(\underline{s})-F_1(\underline{s}))$, and find $M\in\mathbb{N}^+$ such that for any $m\geq M$, $F_1(\underline{s}_*^{i_m})-F_1(\underline{s})<\epsilon$.

For any $m\geq M$, denote a random variable $Z(i_m)=1\{i_m\text{ chooses $0$ without any observation}\}$. We know that $Z(i_M),Z(i_{M+1}),\cdots$ are mutually independent, and that for any $m\geq M$, $Z(i_m)$ is equal to 1 with probability greater than $F_0(\underline{s})$ when $\theta=0$ and with probability less than $F_1(\underline{s})+\epsilon=\frac{1}{2}(F_0(\underline{s})+F_1(\underline{s}))<F_0(\underline{s})$ when $\theta=1$.

By the weak law of large numbers, for any $\rho>0$, we have
\begin{align*}
\lim_{n\rightarrow \infty}Prob(\frac{\sum_{m=M}^{M+n}Z(i_m)}{n+1}<F_0(\underline{s})-\rho|\theta=0)&=0\\
\lim_{n\rightarrow \infty}Prob(\frac{\sum_{m=M}^{M+n}Z(i_m)}{n+1}>\frac{1}{2}(F_0(\underline{s})+F_1(\underline{s}))+\rho|\theta=1)&=0.
\end{align*}
Take $\rho<\frac{1}{2}(F_0(\underline{s})-F_1(\underline{s}))$, and consider the following sub-optimal strategy for any agent $i_m$, $m>M$: observe the neighborhood $\{M,M+1,\cdots,M+n\}$ where $n=\min\{m-1-M,K(i_m)-1\}$. Choose $0$ if $\frac{\sum_{m=M}^{M+n}Z(i_m)}{n+1}\geq F_0(\underline{s})-\rho$, and $1$ otherwise. By the above two conditions, we know that the probability of making the right choice converges to $1$ under either state when the agent's number of observations approaches infinity. Hence, maximal learning occurs.
\end{proof}

\begin{proof}[Proof of Proposition 9]
For all $i\in\{1,\cdots,M\}$, let $\hat{s}^i$ denote the positive private signal that will make an agent indifferent between her private beliefs and the most extreme belief induced by $C_i$, i.e., $\hat{s}^i$ is characterized by
\begin{align*}
\max_{y_k\text{ for all }k\in C_i}\mathcal{P}_{\sigma^*}(\theta=0|a_k=y_k\text{ for all }k\in C_i)=\frac{f_1(\hat{s}^i)}{f_1(\hat{s}^i)+f_0(\hat{s}^i)}.
\end{align*}
Let $\hat{s}=\max_i \hat{s}_i<1$. Let $X_n$ denote the event that agent $n$ can only observe within some $C_i$, $i\in\{1,\cdots,M\}$; let $Y_n$ denote the event that agent $n$ can observe within $\{1,\cdots,n-1\}$.

Consider the term $\frac{\mathcal{P}_{\sigma^*}(a_{k+1}=0|h,1)}{\mathcal{P}_{\sigma^*}(a_{k+1}=0|h,0)}$ in the proof of Lemma 3. It can now be written as
\begin{align*}
\frac{\mathcal{P}_{\sigma^*}(a_{k+1}=0|h,1)}{\mathcal{P}_{\sigma^*}(a_{k+1}=0|h,0)}=\frac{\epsilon_n\mathcal{P}_{\sigma^*}(a_{k+1}=0|h,1,X_{k+1})+(1-\epsilon_n)\mathcal{P}_{\sigma^*}(a_{k+1}=0|h,1,Y_{k+1})}{\epsilon_n\mathcal{P}_{\sigma^*}(a_{k+1}=0|h,0,X_{k+1})+(1-\epsilon_n)\mathcal{P}_{\sigma^*}(a_{k+1}=0|h,0,Y_{k+1})}.
\end{align*}
We know that $\mathcal{P}_{\sigma^*}(a_{k+1}=0|h,1,Y_{k+1})\leq \mathcal{P}_{\sigma^*}(a_{k+1}=0|h,0,Y_{k+1})$. Thus, we have
\begin{align*}
\frac{\mathcal{P}_{\sigma^*}(a_{k+1}=0|h,1)}{\mathcal{P}_{\sigma^*}(a_{k+1}=0|h,0)}&\leq\frac{\epsilon_n\mathcal{P}_{\sigma^*}(a_{k+1}=0|h,1,X_{k+1})+(1-\epsilon_n)\mathcal{P}_{\sigma^*}(a_{k+1}=0|h,0,Y_{k+1})}{\epsilon_n\mathcal{P}_{\sigma^*}(a_{k+1}=0|h,0,X_{k+1})+(1-\epsilon_n)\mathcal{P}_{\sigma^*}(a_{k+1}=0|h,0,Y_{k+1})}\\
&\leq\frac{\epsilon_n\mathcal{P}_{\sigma^*}(a_{k+1}=0|h,1,X_{k+1})+(1-\epsilon_n)}{\epsilon_n\mathcal{P}_{\sigma^*}(a_{k+1}=0|h,0,X_{k+1})+(1-\epsilon_n)}.
\end{align*}

By an argument similar to that in the proof of Lemma 3, we also have
\begin{align*}
\mathcal{P}_{\sigma^*}(a_{k+1}=0,\text{ observe}|h,1,X_{k+1})\leq\mathcal{P}_{\sigma^*}(a_{k+1}=0,\text{ observe}|h,0,X_{k+1})
\end{align*}
which implies that
\begin{align*}
&\frac{\mathcal{P}_{\sigma^*}(a_{k+1}=0|h,1)}{\mathcal{P}_{\sigma^*}(a_{k+1}=0|h,0)}\\
\leq&\frac{\epsilon_n(\mathcal{P}_{\sigma^*}(a_{k+1}=0,\text{ not observe}|h,1,X_{k+1})+\mathcal{P}_{\sigma^*}(a_{k+1}=0,\text{ observe}|h,0,X_{k+1}))+(1-\epsilon_n)}{\epsilon_n(\mathcal{P}_{\sigma^*}(a_{k+1}=0,\text{ not observe}|h,0,X_{k+1})+\mathcal{P}_{\sigma^*}(a_{k+1}=0,\text{ observe}|h,0,X_{k+1}))+(1-\epsilon_n)}\\
\leq&\frac{\epsilon_n F_1(\hat{s})+(1-\epsilon_n)}{\epsilon_n F_0(\hat{s})+(1-\epsilon_n)}\\
\leq&\frac{\epsilon F_1(\hat{s})+(1-\epsilon)}{\epsilon F_0(\hat{s})+(1-\epsilon)}<1.
\end{align*}

The argument for Theorem 2 can then be applied at this juncture to prove that if $\lim_{n\rightarrow\infty}K(n)=\infty$, $\lim_{n\rightarrow\infty}\mathcal{P}_{\sigma^*}(a_n=\theta|Y_n)=1$. Therefore, we have
\begin{align*}
\lim\inf_{n\rightarrow\infty} \mathcal{P}_{\sigma^*}(a_n=\theta)\geq (1-\epsilon)\lim_{n\rightarrow\infty}\mathcal{P}_{\sigma^*}(a_n=\theta|Y_n)=1-\epsilon.
\end{align*}
Hence, $\epsilon$-maximal learning occurs.
\end{proof}

\newpage
\bibliographystyle{acm}
\bibliography{reference}

\end{document}